\documentclass[12pt, draftclsnofoot, onecolumn]{IEEEtran}
\usepackage[latin9]{inputenc}
\usepackage{array}
\usepackage{float}
\usepackage{mathtools}
\usepackage{amsmath}
\usepackage{amsthm}
\usepackage{amssymb}
\usepackage{graphicx}
\usepackage{wasysym}
\usepackage{setspace}
\usepackage{color}
\usepackage{bm}
\usepackage{cases}
\usepackage[top=1.9cm, left=1.9cm, bottom=1.9cm, right=1.9cm]{geometry}
\usepackage{tikz}
\usepackage{flowchart}
\usetikzlibrary{matrix,shapes,arrows,positioning,chains}

\makeatletter

\providecommand{\tabularnewline}{\\}

\floatstyle{ruled}
\newfloat{algorithm}{tbp}{loa}
\providecommand{\algorithmname}{Algorithm}
\floatname{algorithm}{\protect\algorithmname}

\theoremstyle{plain}
\newtheorem{thm}{\protect\theoremname}
\theoremstyle{remark}

\theoremstyle{plain}

\theoremstyle{plain}
\newtheorem{lem}[thm]{\protect\lemmaname}

\usepackage{epsfig}
\usepackage{subfigure}
\usepackage{cite}
\usepackage{graphicx}
\usepackage{multirow}
\usepackage{array}
\usepackage{enumerate}
\usepackage{graphicx}
\usepackage{times}
\usepackage{algorithm}
\usepackage{algorithmic}
\newtheorem{theorem}{Theorem}

\newtheorem{proposition}{Proposition}
\newtheorem{definition}{Definition}
\DeclareMathOperator*{\argmin}{argmin}
\DeclareMathOperator*{\argmax}{argmax}
\makeatother

\providecommand{\lemmaname}{Lemma}
\providecommand{\propositionname}{Proposition}
\providecommand{\remarkname}{Remark}
\providecommand{\theoremname}{Theorem}
\providecommand{\theoremname}{Definition}

\begin{document}

\title{Resource Allocation  and Power Control in Cooperative Small Cell Networks in Frequency Selective Channels with Backhaul Constraint}

\author{Jonggyu Jang,~\IEEEmembership{Student Member,~IEEE}, Hyun~Jong~Yang,~\IEEEmembership{Member,~IEEE}, and Hyekyung
Jwa,~\IEEEmembership{Member,~IEEE}
\thanks{
This work has been submitted
to the IEEE for possible publication. Copyright may be transferred
without notice, after which this version may no longer be accessible.

J. Jang and H. J. Yang (corresponding author) are with the School of Electrical and Computer Engineering, Ulsan National Institute of Science and Technology (UNIST), Ulsan 44919, Republic of Korea, (e-mail: \{jonggyu,
hjyang\}@unist.ac.kr). H. J. Yang is also with EgoVid Inc, Ulsan 44919, Republic of Korea.

H. Jwa is with the Wireless Application Research Department of, Electronics and Telecommunications Research Institution (ETRI), Daejeon 34129, Republic of Korea, (email: hkjwa@etri.re.kr).}
}
\maketitle

\begin{abstract}
A joint resource allocation (RA), user association (UA), and power control (PC) problem is addressed for proportional fairness maximization in a cooperative multiuser downlink small cell network
with limited backhaul capacity, based on orthogonal frequency division multiplexing. Previous studies have relaxed the per-resource-block (RB) RA and UA problem to a continuous optimisation problem based on long-term signal-to-noise-ratio, because the original problem is known as a combinatorial NP-hard problem. We tackle the original per-RB RA and UA problem to obtain a near-optimal solution with feasible complexity. We show that the conventional dual problem approach for RA cannot find the solution satisfying the conventional KKT conditions. Inspired by the dual problem approach, however, we derive the first order optimality conditions for the considered RA, UA, and PC problem, and propose a sequential optimization method for finding the solution.  The overall proposed scheme can be implemented with feasible complexity even with a large number of system parameters. Numerical results show that the proposed scheme achieves the proportional fairness close to its outer bound with unlimited backhaul capacity in the low backhaul capacity regime and to that of a carefully-designed genetic algorithm with excessive generations but without backhaul constraint in the high backhaul capacity regime.
\end{abstract}


\section{Introduction}

It has long been a challenge to suppress intercell interference in cellular mobile networks, and thereby improving the system throughput and spectral efficiency. Indeed, it is known that the cell densification may degrade the sum-rate unless the level of intercell interference is kept low enough compared to the desired signal \cite{A_Gupta_WCL15}. The concept of ``small cells'' is one of the enablers of the next generation mobile network requiring extremely high data rate connections, where multiple small cell base stations (SBSs) in proximity are clustered to make a hotspot area providing high data rate connectivity. As the SBS cell size becomes smaller to further increase the sum-rate, user association (UA), resource allocation (RA), and power control (PC) should be carefully designed to mitigate intercell interference, particularly for cell-edge users. The optimisation for UA, RA, and PC can be considered to maximize the fairness of the users \cite{M_T.Kawser_JEE12} or the sum-capacity of the total system \cite{Q_Kuang16_TSP, H_ju13_WCL, S_Han_PIMRC12, Z_shen_GLC03}. The focus of this paper is on the proportional fairness maximization, where the aim is to maximize the geometric mean of users' rates, compromising between the fairness and sum-capacity maximization.

A variety of literature have been around to tackle the joint optimisation problem of UA, RA, and PC for heterogenous multicell networks based on orthogonal frequency division multiplexing (OFDM) \cite{S_Han_PIMRC12,Y_Zhang_TWC04,N_Ksairi_TSP10, N_Ksairi_TSP10_2}. In SBS networks, a reliable direct connection, i.e., X2 interface in LTE-A, is assumed between SBSs within a cluster, and hence the aforementioned problem can be solved cooperatively across the SBSs in a cluster, allowing the exchange of the channel gain values among the SBSs in the cluster.

A rich body of the literature assumes unique BS association \cite{S_Han_PIMRC12,Y_Zhang_TWC04,N_Ksairi_TSP10, N_Ksairi_TSP10_2,D_Fooladivanda13_TWC,Derrick_TWC,Ghime_JSAC,Kaiming_JSAC,Q_Han15,Q_Ye_TWC,T_Bu_INFO,T_Girici_JCN,Y.C.Chen}, in which a user can only be served by a unique BS, i.e., the user can receive data only from a single BS out of all the BSs\footnote{Note that although the UA, RA, and PC can be optimized cooperatively across multiple SBSs, every user is associated with a unique BS for data reception.}. In frequency-selective fading channels, the RA and UA needs to be done on a per-resource-block (RB) or per-subcarrier basis, which generally results in a combinatorial NP-hard problem requiring exponential computational complexity with respect to the number of RBs and/or users to find the optimal solution. Several efforts have been made to relax this combinatorial problem to more tractable problems,
i) reducing search space heuristically
\cite{S_Han_PIMRC12, Derrick_TWC},
ii) simply relaxing  integer variables to real-valued variables
\cite{Q_Ye_TWC, A_Abdelnasser15_TCM}, or iii) only using SNR avreged over all the RBs for each BS \cite{Q_Ye_TWC,D_Fooladivanda13_TWC,Q_Han15,Kaiming_JSAC, T_Bu_INFO, K_son_TWC}. With the SNR averaged out for all the RBs, ignoring frequency selectivity, the RA problem is to find the proportion of the total bandwidth, allocated to each user, thus yielding a continuous optimisation problem with real-valued variables. In particular, the authors of \cite{Q_Ye_TWC} found a near-optimal solution for the UA and RA for the frequency non-selective fading case.

On the other hand, a user can be jointly associated with multiple SBSs within a cluster to further improve  the system throughput \cite{Q_Ye_TWC,S.Maghsudi_16,The_Future_of_WN,D.Bethanabhotla_14,Z_Cui14_CISS, Z_shen_GLC03,DBLP_17}, in which a user can be simultaneously served via different RBs of different SBSs. Although the total throughput can be significantly enhanced using the multiple-BS association, a careful consideration is needed for the total amount of downlink data to be transmitted by each SBS not to exceed the backhaul capacity. In particular, non-ideal backhaul such as wireless backhaul with highly limited capacity is the highest priority of service operators of LTE-A \cite{3gpp_36_872}.

In this paper, our focus is on multiple-BS association with limited backhaul capacity for frequency-selective fading channels. In particular, the per-RB UA, RA, and PC problem is tackled in pursuit of maximizing the proportional fairness.

\subsection{Related Works}
The authors of \cite{Q_Ye_TWC} formulate and solve a real-valued convex problem of joint UA and RA assuming frequency non-selective fading with fixed power. In \cite{D.Bethanabhotla_14,DBLP_17}, a similar technique was used for UA and load balancing in the multicell frequency non-selective fading massive multi-input multi-output channel. Distributed UA schemes for given resource element (RE) with the SBSs harvesting energy are proposed in \cite{S.Maghsudi_16}, where each user selects a SBS based on the statistical analysis of the amount of energy harvesting and energy consumption of the SBSs. These studies, however, assume frequency non-selective fading channels with ideal  backhaul capacity, thus yielding a limited application in practical environment.

In \cite{Z_Cui14_CISS}, the joint problem of UA, RA, and PC is tackled with limited backhaul capacity constraint, where multiple-BS association is allowed. In particular, the authors of \cite{Z_Cui14_CISS} merged the integer variables for UA and RA into the real-valued power variables, thereby yielding a continuous optimisation problem. However, this study also assumes frequency non-selective fading, where the problem is formulated on a long-term basis based only on long-term SNR. Therefore the backhaul constraint cannot be satisfied at all time, particularly if channel gain is temporarily high for all the REs at random.

To the best of the authors' knowledge, in spite of its importance, the joint problem of UA, RA, and PC has never been solved with feasible computational complexity in frequency selective channels, where the RA, UA, and PC are carried out on a per-RB basis.

\subsection{Contribution}
We formulate the joint problem of UA,  RA, and PC for frequency selective fading channels with limited backhaul capacity, and show that the well-known conventional dual problem approach cannot be used for the considered problem. Inspired by the dual problem approach, however, we derive the first order optimality conditions, and then propose a two-step cascaded algorithm to find the solutions of the UA, RA, and PC problems sequentially with feasible computational complexity. In the proposed UA and RA algorithm, the gap between the 2-distance ring points, i.e., local optimal point, and the solution of the proposed algorithm is derived in terms of the lagrange variables, which asymptotically vanishes as the number of variables increases. For the PC problem, a zero-sum-game approach for the power allocated to the users of each SBS is proposed with sum-power constraint based on the first-order optimality condition.

Simulation results show that the proposed scheme exhibits the proportional fairness performance close to the outer bound with unlimited power assumption in the low backhaul capacity regime and to that of the genetic algorithm under unlimited backhaul capacity, which generally finds a near-global optimal solution with excessive generations, in the high backhaul capacity regime.

\subsection{Organization of this paper}

The remainder of this paper is organized as follow. Section \ref{sec:System-Model}
introduces the system model and formulates the problem. Section \ref{sec:dual-problem} analyzes the dual problem approach.
Section \ref{sec:prop} presents the proposed UA, RA, and PC algorithms to find the solution.
Section \ref{sec:Numerical-Results} provides numerical results, and  Section \ref{sec:Conclusions} concludes the paper.

\section{System Model and Problem Formulation \label{sec:System-Model}}

We consider a downlink SBS cluster composed of $J$ SBSs and $N$ users based on  orthogonal frequency
division multiplexing (OFDM). Assuming separate frequency carrier for the macro-cell BSs, e.g., Scenario 2a of the 3GPP small cell scenarios \cite{3gpp_36_872}, there is no interference from the macro-cell BSs. Assuming frequency reuse 1 for the SBSs in the cluster, inter-cluster interference is neglected, which is dominated by intra-cluster interference. The UA and RA is carried out cooperatively across all the SBSs only within the cluster, assuming the exchange of per-RB channel gain information among SBSs via a direct interface, such as X2 interface in LTE-A.
Each SBS is connected to the core network via backhaul link with limited capacity. The total bandwidth is divided into $C$ frequency-division
RBs, each of which is a group of multiple or single subcarriers. In frequency-selective fading, each RB has different channel gain for each user.

Let $h_{ij}^{(c)}$ denote the channel gain from SBS $j$ to user $i$ on RB $c$, where $j\in\mbox{\ensuremath{\mathcal{B}}}=\{1,\,2,\,\cdots,\,J\}$,
$i\in\mbox{\ensuremath{\mathcal{N}}}=\{1,\,2,\,\cdots,\,N\},$ and
$c\in\mbox{\ensuremath{\mathcal{C}}}=\{1,\,2,\,\cdots,\,C\}$. Assuming quasi-static block fading, i.e., $h_{ij}^{(c)}$ is constant for a frame and changes to the next value randomly, all the SBSs in the cluster share the channel gain values for all the users within the cluster. This global CSI assumption within a cluster is feasible, because high data rate interface, such as optical fiber, between the SBSs in the same cluster is considered with high priority in the commercialized network \cite{3gpp_36_872}.
The signal-to-interference-plus-noise ratio (SINR) when SBS $j$
serves user $i$ on RB $c$ is denoted as
\begin{equation}
\mathrm{SINR}_{ij}^{(c)}=\frac{\left|h_{ij}^{(c)}\right|^{2}p_{j}^{(c)}}{\sigma^{2}+\sum_{k\neq j, k\in \mathcal{B}}\left|h_{ik}^{(c)}\right|^{2}p_{k}^{(c)}},
\end{equation}
where $\sigma^{2}$ represents the variance of additive white Gaussian
noise (AWGN), and $p_{j}^{(c)}$ is the transmission
power of SBS $j$ on RB $c$, constrained by
\begin{equation}
p_{j}^{(c)}\geq0,\,\,\forall j\in\mathcal{B},c\in\mathcal{C},\label{eq:Constraint_power_nonnegative}
\end{equation}
\begin{equation}
\sum_{c\in\mathcal{C}}p_{j}^{(c)}\leq P_{j,\textrm{max}},\,\,\forall j\in\mathcal{B},\label{eq:Constraint_power_sumpower}
\end{equation}
where $P_{j,\textrm{max}}$ denotes the maximum total transmission power of SBS $j$ across
all the frequency blocks. The data rate of user $i$ served by
SBS $j$ on RB $c$ is represented by
\begin{equation}
R_{ij}^{(c)}=W\log_{2}(1+\mathrm{SINR}_{ij}^{(c)}),
\end{equation}
where $W$ denotes the bandwidth of each RB. We define a binary variable $x_{ij}^{(c)}$  to represent UA and RA as follow:
\begin{equation}
x_{ij}^{(c)}=\begin{cases}
1, & \text{if user $i$\,\ is served by SBS $j$\,\ on RB $c$,}\\
0, & \text{otherwise.}
\end{cases}\label{eq:1or0}
\end{equation}
We assume up to one user
can be served on each RB. Note that each user can be served by different RBs of different SBSs, i.e., joint SBS association is allowed. These two conditions are denoted as
\begin{equation}
\sum_{i\in\mathcal{N}}x_{ij}^{(c)}=1,\forall j\in\mathcal{B},\,c\in\mathcal{C}.\label{eq:sum=00003D1}
\end{equation}
Each transmission of SBSs is constrained by backhaul capacity. The constraint for backhaul capacity is denoted as
\begin{equation}
\sum_{c\in\mathcal{C}}\sum_{i\in\mathcal{N}}R_{ij}^{(c)}x_{ij}^{(c)}\leq Z_{j},\forall j\in\mathcal{B},\label{eq:condition_bh}
\end{equation}
where $Z_{j}$ denotes the
backhaul capacity of SBS $j$.

The proportional fairness of the network with given UA and RA is denoted as
\begin{equation} \label{eq:prop_fairness}
U(\mathbf{X},\mathbf{P})=\sum_{i\in\mathcal{N}}\log\left(\sum_{j\in\mathcal{B}}\sum_{c\in\mathcal{C}}R_{ij}^{(c)}x_{ij}^{(c)}\right),
\end{equation}
where $\mathbf{X}\in \{0,1\}^{N \times B\times C}$ with $[\mathbf{X}]_{ijc}=x_{ij}^{(c)}$ and $\mathbf{P}\in \mathbb{R}^{B\times C}$ with $[\mathbf{P}]_{jc}=p_{j}^{(c)}$.
The goal is to find the solution for $\mathbf{X}$ and $\mathbf{P}$ that maximize the proportional fairness under given backhaul constraint. The mixed integer optimisation problem is formulated as
\begin{subequations} \label{eq:P1}
\begin{align}
\mathrm{P1: } \max_{\mathbf{X}, \mathbf{P}} & \sum_{i\in\mathcal{N}}\log\left(\sum_{j\in\mathcal{B}}\sum_{c\in\mathcal{C}}R_{ij}^{(c)}x_{ij}^{(c)}\right) \\
\textrm{s.t.  } & \sum_{i\in\mathcal{N}} x_{ij}^{(c)}=1,\forall j\in\mathcal{B},\forall c\in\mathcal{C}\\
 & \sum_{c\in\mathcal{C}}\sum_{i\in\mathcal{N}}R_{ij}^{(c)}x_{ij}^{(c)}\leq Z_{j},\forall j\in\mathcal{B}\\
 & \sum_{c\in\mathcal{C}}p_{j}^{(c)}\leq P_{j,\textrm{max}},\forall j\in\mathcal{B}\\
 & R_{ij}^{(c)}=W\log_{2}\left(1+\frac{\left|h_{ij}^{(c)}\right|^{2}p_{j}^{(c)}}{\sigma^{2}+\sum_{k\neq j}\left|h_{ik}^{(c)}\right|^{2}p_{k}^{(c)}}\right),\forall i\in\mathcal{N},\forall j\in\mathcal{B},\forall c\in\mathcal{C}\\
 & x_{ij}^{(c)}\in\{0,1\},\forall i\in\mathcal{N},\forall j\in\mathcal{B},\forall c\in\mathcal{C}\\
 & p_{j}^{(c)}\geq0,\forall j\in\mathcal{B},\forall c\in\mathcal{C}
\end{align}
\end{subequations}
Because the joint optimisation problem \eqref{eq:P1} is not tractable for both $\mathbf{X}$ and $\mathbf{P}$, we decompose
the problem to solve $\mathbf{X}$ and $\mathbf{P}$ sequentially
in following Sections.

\section{Duality Analysis of UA and RA\label{sec:dual-problem}}
In this section, $\mathbf{X}$, which denotes UA and RA, is obtained with given transmission power $\mathbf{P}$ and
data rate $R_{ij}^{(c)}$. The joint UA
and RA problem is formulated with given $\mathbf{P}$ from problem \eqref{eq:P1}.
Then, the problem is formulated by adding the variable $\lambda_{i}$ without losing any optimality:
\begin{subequations}\label{eq:P_UA}
\begin{align}
\mathrm{P2: } \max_{\mathbf{X}} & \sum_{i\in\mathcal{N}}\log\lambda_{i}\\
\textrm{s.t.  } & \lambda_{i}=\sum_{j\in\mathcal{B}}\sum_{c\in\mathcal{C}}R_{ij}^{(c)}x_{ij}^{(c)}, \forall i\in\mathcal{N} \label{eq:lambda_ineq}\\
& \sum_{i\in\mathcal{N}}x_{ij}^{(c)}=1, \forall j\in\mathcal{B},\forall c\in\mathcal{C}\label{eq:P2_c}\\
& \sum_{c\in\mathcal{C}}\sum_{i\in\mathcal{N}} R_{ij}^{(c)}x_{ij}^{(c)}\leq Z_j, \forall j\in\mathcal{B}\label{eq:backhaul_ineq}\\
& x_{ij}^{(c)}\in\{0,1\},\forall i\in\mathcal{N},\forall j\in\mathcal{B},\forall c\in\mathcal{C}\label{eq:P2_e}
\end{align}
\end{subequations}
The lagrangian expression of the problem \eqref{eq:P_UA} except constraints \eqref{eq:P2_c} and \eqref{eq:P2_e} is denoted as
\begin{align} \label{eq:Lag}
L(\mathbf{X}, \bm{\lambda}, \bm{\mu}, \bm{\nu})  =\sum_{i\in\mathcal{N}}\log\lambda_{i}+\sum_{i\in\mathcal{N}}\mu_{i}\left(\sum_{j\in\mathcal{B}}\sum_{c\in\mathcal{C}}R_{ij}^{(c)}x_{ij}^{(c)}-\lambda_{i}\right)
 +\sum_{j\in\mathcal{B}}\nu_{j}\left(Z_{j}-\sum_{i\in\mathcal{N}}\sum_{c\in\mathcal{C}}R_{ij}^{(c)}x_{ij}^{(c)}\right),
\end{align}
where $\mu_{i}$ and $\nu_{j}$ are the lagrange multipliers corresponding to the constraints  \eqref{eq:lambda_ineq} and \eqref{eq:backhaul_ineq},  $\bm{\mu}\in\mathbb{R}^{N}$ with $[\bm{\mu}]_{i}=\mu_{i}$ , $\bm{\nu}\in\mathbb{R}_{+}^{B}$ with $[\bm{\nu}]_{j}=\nu_{j}$, and $\bm{\lambda}\in\mathbb{R}^{N}$ with $[\bm{\lambda}]_{i}=\lambda_{i}$. Then, the dual function of \eqref{eq:P_UA}
is given by
\begin{equation} \label{eq:dual_define}
g(\bm{\mu}, \bm{\nu})=\sup_{\mathbf{X}\in\mathcal{X}_{f},\bm{\lambda}}L(\mathbf{X},\bm{\lambda},\bm{\mu},\bm{\nu}),
\end{equation}
where $\mathcal{X}_{f}$ represents the domain of $\mathbf{X}$ that satisfies the constraints \eqref{eq:P2_c} and \eqref{eq:P2_e}, defined as
\begin{equation} \label{eq:X_f}
    \mathcal{X}_f = \left\{ \mathbf{X}:  \sum_{i\in\mathcal{N}}x_{ij}^{(c)}=1, \forall(j,c), \hspace{10pt} x_{ij}^{(c)}\in\{0,1\},\forall (i,j,c) \right\}.
\end{equation}
The aim first is to solve $\sup_{\mathbf{X}\in\mathcal{X}_{f},\bm{\lambda}}L(\mathbf{X},\bm{\lambda},\bm{\mu},\bm{\nu})$ in \eqref{eq:dual_define}.
To this end, a problem, which obatins $g(\bm{\mu}, \bm{\nu})$, is defined as follow:
\begin{subequations} \label{eq:x_lambda_prob}
\begin{align}
\max_{\mathbf{X}, \bm{\lambda}} & L(\mathbf{X}, \bm{\lambda}, \bm{\mu}, \bm{\nu})\\
\textrm{s.t. } & \sum_{i\in\mathcal{N}}x_{ij}^{(c)}=1, \forall j\in\mathcal{B},\forall c\in\mathcal{C} \label{eq:cond_x_sum}\\
& x_{ij}^{(c)}\in\{0,1\}, \forall i\in\mathcal{N},\forall j\in\mathcal{B},\forall c\in\mathcal{C} \label{eq:cond_x_binary}
\end{align}
\end{subequations}
An optimal value $\bm{\lambda}^{*}$ maximizing the lagrangian \eqref{eq:Lag}
is obtained by partial derivative with respect to $\lambda_{i}$ as follow:
\begin{equation}
\frac{\partial L(\mathbf{X}, \bm{\lambda}, \bm{\mu}, \bm{\nu})}{\partial\lambda_{i}}=\frac{1}{\lambda_{i}}-\mu_{i}=0
\longrightarrow \lambda_{i} =  \frac{1}{\mu_{i}}. \label{eq:lambda}
\end{equation}
Inserting \eqref{eq:lambda} into the lagrangian \eqref{eq:Lag} gives us
\begin{equation}\label{eq:L_given_lambda}
\tilde{L}(\mathbf{X}, \bm{\mu}, \bm{\nu}) = \sum_{i\in\mathcal{N}}\log\frac{1}{\mu_{i}}
+\sum_{j\in\mathcal{B}}\sum_{c\in\mathcal{C}}\sum_{i\in\mathcal{N}}R_{ij}^{(c)}x_{ij}^{(c)}\left(\mu_{i}-\nu_{j}\right)
+\sum_{j\in\mathcal{B}}\nu_{j}Z_{j}-|\mathcal{N}|.
\end{equation}
Now, the aim is to maximize $\tilde{L}(\mathbf{X}, \bm{\mu}, \bm{\nu})$ over $\mathbf{X}$. Since $\mathbf{X}$ should satisfy the constraints \eqref{eq:cond_x_sum} and  \eqref{eq:cond_x_binary}, the optimal $\mathbf{X}$ is obtained by
\begin{equation}\label{eq:x_Alloc}
{x_{ij}^{(c)}}=F_{ij}^{(c)}(\bm{\mu},\bm{\nu}),
~\textrm{ where }
F_{ij}^{(c)}(\bm{\mu},\bm{\nu})=\begin{cases}
1, & \text{if } i=\argmax_{\hat{i}\in\mathcal{N}}R_{\hat{i}j}^{(c)}\left(\mu_{\hat{i}}-\nu_{j}\right),\\
0, & \text{otherwise.}
\end{cases}
\end{equation}
By substituting \eqref{eq:x_Alloc} into \eqref{eq:L_given_lambda}, the dual function $g(\cdot)$ for given $\boldsymbol{\mu}$ and $\boldsymbol{\nu}$ is represented as
\begin{align}\label{eq:dual function}
   g(\bm{\mu},\bm{\nu}) =\sum_{i\in\mathcal{N}}\log{\frac{1}{\mu_{i}}}+
    \sum_{j\in\mathcal{B}}\sum_{c\in\mathcal{C}}\max_{i\in\mathcal{N}} R_{ij}^{(c)}\left(\mu_{i}-\nu_{j}\right)+\sum_{j\in\mathcal{B}}\nu_{j}Z_{j}-|\mathcal{N}|.
\end{align}
The dual problem for $\bm{\mu}$ and $\bm{\nu}$ then is represented as follow:
\begin{subequations}  \label{eq:min_g_mu_nu}
\begin{align}
\min_{\bm{\mu},\bm{\nu}} \,\,\, & g(\bm{\mu},\bm{\nu})\\
\textrm{s.t. } & \nu_{j}\geq0,\,\forall j\in\mathcal{B}
\end{align}
\end{subequations}
At this point, let us denote the optimal solution of $(\bm{\mu},\bm{\nu})$ of the problem \eqref{eq:min_g_mu_nu} as $\bm{\mu}^*\in\mathbb{R}^N$ with $[\bm{\mu}^*]_i=\mu_i^*$ and $\bm{\nu}^*\in\mathbb{R}_+^{J}$ with $[\bm{\nu}^*]_j=\nu_j^*$. In addition, from \eqref{eq:lambda} and \eqref{eq:x_Alloc}, we define $\mathbf{X}^*$ and $\bm{\lambda}^*$ as
\begin{equation} \label{eq:x_star}
\begin{cases}
{x_{ij}^{(c)}}^* = F_{ij}^{(c)}(\bm{\mu}^*,\bm{\nu}^*),\\
 \lambda_i^* = \frac{1}{\mu_i^*},
\end{cases}
\end{equation}
where $\mathbf{X}^*\in\{0,1\}^{N\times B\times C}$ with $[\mathbf{X}^*]_{ij}={x_{ij}^{(c)}}^*$ and $\bm{\lambda}^*\in\mathbb{R}^{N}$ with $[\bm{\lambda}^*]_j=\lambda_i^*$. That is, $(\mathbf{X}^*, \bm{\lambda}^*,\bm{\mu}^*, \bm{\nu}^*)$ is the dual solution of the problem \eqref{eq:P_UA}.

 A closed-form solution of the problem (\ref{eq:min_g_mu_nu}) is difficult to obtain, since $g(\bm{\mu},\bm{\nu})$ is a convex function but may be non-differentiable for $\bm{\mu}^*$ and $\bm{\nu}^*$ as shown in the following lemmas.
\begin{proposition}\label{lem:1}
$g(\bm{\mu}, \bm{\nu})$ is a convex function with respect to $\boldsymbol{\mu}$ and $\boldsymbol{\nu}$.
\end{proposition}
\begin{IEEEproof}
The Hessian of the first term of $g(\bm{\mu}, \bm{\nu})$ in \eqref{eq:dual function} is semi-positive definite, since it is a diagonal matrix with all positive  diagonal elements. The second term of $g(\bm{\mu}, \bm{\nu})$ is also convex, since the maximum of the affine functions is convex. Therefore, the dual function $g(\bm{\mu}, \bm{\nu})$ is convex respect to $\bm{\mu}$ and $\bm{\nu}$.
\end{IEEEproof}
\begin{theorem}\label{thm:global}
The solution $(\boldsymbol{\mu}^*,\boldsymbol{\nu}^*)$ of the dual problem \eqref{eq:min_g_mu_nu} does not satisfy KKT condition if and only if there exist $j\in \mathcal{B}$ such that $\nu_j^*>0$, or $\boldsymbol{\nu}^*=\boldsymbol{0}$ and  $N<\sum_{j\in\mathcal{B}}\sum_{c\in\mathcal{C}}\max_{i\in\mathcal{N}}\frac{R_{ij}^{(c)}}{\sum_{k\in\mathcal{B}}\sum_{l\in\mathcal{C}}R_{ik}^{(l)}{x_{ik}^{(l)}}^{*}}$. In other words, KKT conditions are satisfied at $(\boldsymbol{\mu}^*,\boldsymbol{\nu}^*)$ only if $\boldsymbol{\nu}^*=\boldsymbol{0}$ and  $N\ge\sum_{j\in\mathcal{B}}\sum_{c\in\mathcal{C}}\max_{i\in\mathcal{N}}\frac{R_{ij}^{(c)}}{\sum_{k\in\mathcal{B}}\sum_{l\in\mathcal{C}}R_{ik}^{(l)}{x_{ik}^{(l)}}^{*}}$.
\end{theorem}
\begin{IEEEproof}
The KKT condition of the problem \eqref{eq:min_g_mu_nu} is denoted as
\begin{equation}\label{eq:KKT_condition_b}
    \begin{cases}
    \frac{\partial g(\bm{\mu},\bm{\nu})}{\partial\nu_j}\geq0,&\forall j\in\mathcal{B}\\
    \nu_j\frac{\partial g(\bm{\mu},\bm{\nu})}{\partial\nu_j}=0, &\forall j\in\mathcal{B}\\
    \frac{\partial g(\bm{\mu},\bm{\nu})}{\partial\mu_i}=0,  &\forall i\in\mathcal{N}
    \end{cases}
\end{equation}
i) $\nu_j^{*}>0$ for any $j\in\mathcal{B}$: \\
Let us assume that $\nu_{\hat{j}}^*>0$ for some $\hat{j}\in \mathcal{B}$. Because $R_{ij}^{(c)}$, $i=1,\ldots, N$, $j=1, \ldots, B$, $c=1, \ldots, C$, are continuous random variables, for any given binary matrix $\mathbf{X}$ and real-valued constant $Z_j$, we can have  $Z_j=\sum_{i\in\mathcal{N}}\sum_{c\in\mathcal{C}}R_{ij}^{(c)}x_{ij}^{(c)}$ with probability 0. Since inserting $(\boldsymbol{\mu}^*,\boldsymbol{\nu}^*)$ into \eqref{eq:dual function} gives us
\begin{equation}\label{eq:g_star}
    g(\boldsymbol{\mu}^*,\boldsymbol{\nu}^*) =\sum_{i\in\mathcal{N}}\log{\frac{1}{\mu_{i}^*}}+
    \sum_{j\in\mathcal{B}}\sum_{c\in\mathcal{C}}\sum_{i\in\mathcal{N}} R_{ij}^{(c)}{x_{ij}^{(c)}}^*\left(\mu_{i}^*-\nu_{j}^*\right)+\sum_{j\in\mathcal{B}}\nu_{j}^*Z_{j}-|\mathcal{N}|,
\end{equation}
we have the following with probability $1$:
\begin{equation} \label{eq:dg_dnu}
    \left.\frac{\partial g(\bm{\mu},\bm{\nu})}{\partial\nu_j} \right|_{(\boldsymbol{\mu}^*,\boldsymbol{\nu}^*)} = Z_j-\sum_{i\in\mathcal{N}}\sum_{c\in\mathcal{C}}R_{ij}^{(c)}x_{ij}^{(c)*}\neq0, \forall j\in\mathcal{B}.
\end{equation}
From \eqref{eq:dg_dnu} and $\nu_{\hat{j}}^*>0$, we have
\begin{equation}
    \nu_{\hat{j}}^* \left.\frac{\partial g(\bm{\mu},\bm{\nu})}{\partial\nu_j} \right|_{(\boldsymbol{\mu}^*,\boldsymbol{\nu}^*)}\neq 0,
\end{equation}
which contradicts the second KKT optimality condition in \eqref{eq:KKT_condition_b}.

ii)  $\bm{\nu}^*=\bm{0}$: \\
We shall show that the solution $(\bm{\mu}^*,\bm{\nu}^*)$ of the problem \eqref{eq:min_g_mu_nu} does not satisfies the KKT condition of \eqref{eq:min_g_mu_nu} if $\bm{\nu}^* =\bm{0}$ and $N<\sum_{j\in\mathcal{B}}\sum_{c\in\mathcal{C}}\max_{i\in\mathcal{N}}\frac{R_{ij}^{(c)}}{\sum_{k\in\mathcal{B}}\sum_{l\in\mathcal{C}}R_{ik}^{(l)}{x_{ik}^{(l)}}^{*}}$.
Suppose that $g(\boldsymbol{\mu}^*, \boldsymbol{0})$ satisfies the KKT conditions in \eqref{eq:KKT_condition_b}, then we have
\begin{align} \label{eq:KKT_mu_i}
    \left.\frac{\partial g(\bm{\mu}, \bm{\nu})}{\partial\mu_i}\right|_{(\bm{\mu}^*,\bm{0})}
    &=-\frac{1}{\mu_i^*}+\sum_{j\in\mathcal{C}}\sum_{c\in\mathcal{C}}R_{ij}^{(c)}{x_{ij}^{(c)}}^*=0,
\end{align}
From \eqref{eq:KKT_mu_i}, we have
\begin{equation} \label{eq:mu_star2}
\mu_i^* = \frac{1}{\sum_{j\in\mathcal{C}}\sum_{c\in\mathcal{C}}R_{ij}^{(c)}{x_{ij}^{(c)}}^*},\hspace{10pt}\forall i \in\mathcal{N}.
\end{equation}
Because $\mathbf{X}^*$ is defined as the global maximization function $F_{ij}^{(c)}$ over $\mathcal{X}_f$ as in \eqref{eq:x_Alloc}, we have
\begin{equation}
    g(\bm{\mu}^*, \bm{0}) = \tilde{L}(\mathbf{X}^*,\bm{\mu}^*,\bm{0})\geq \tilde{L}(\mathbf{X},\bm{\mu}^*,\bm{0}), \hspace{10pt} \forall \mathbf{X}\in \mathcal{X}_f,
\end{equation}
which implies that
\begin{equation} \label{eq:inequality1}
        \sum_{i\in\mathcal{N}}\log\frac{1}{\mu_i^*}+\sum_{j\in\mathcal{B}}\sum_{c\in\mathcal{C}}\sum_{i\in\mathcal{N}}R_{ij}^{(c)}x_{ij}^{(c)*}\mu_i^*\geq\sum_{i\in\mathcal{N}}\log\frac{1}{\mu_i^*}+\sum_{j\in\mathcal{B}}\sum_{c\in\mathcal{C}}\sum_{i\in\mathcal{N}}R_{ij}^{(c)}x_{ij}^{(c)}\mu_i^*,\hspace{10pt}\forall \mathbf{X}\in \mathcal{X}_f.
\end{equation}
Inserting \eqref{eq:mu_star2} into \eqref{eq:inequality1} gives us
\begin{equation}\label{eq:condition_N}
    N\geq\sum_{i\in\mathcal{N}}\frac{V_i(\mathbf{X})}{V_i(\mathbf{X}^*)},\hspace{5pt}\forall \mathbf{X} \in \mathcal{X}_f,
\end{equation}
where $V_i(\mathbf{X}) = \sum_{j\in\mathcal{B}}\sum_{c\in\mathcal{C}}R_{ij}^{(c)}x_{ij}^{(c)}$ and  $V_i(\mathbf{X}^*)=\sum_{j\in\mathcal{B}}\sum_{c\in\mathcal{C}}R_{ij}^{(c)}x_{ij}^{(c)*}$.

Therefore, if  $g(\boldsymbol{\mu}^*, \boldsymbol{0})$ satisfies the KKT condition, $N\geq\sum_{i\in\mathcal{N}}\frac{V_i(\mathbf{X})}{V_i(\mathbf{X}^*)},\forall \mathbf{X}\in \mathcal{X}_f$. At this point, the negation of the statement `$N\geq\sum_{i\in\mathcal{N}}\frac{V_i(\mathbf{X})}{V_i(\mathbf{X}^*)},\forall \mathbf{X}\in \mathcal{X}_f$' is given by
\begin{align}
    N & < \max_{\mathbf{X}\in \mathcal{X}_f}\sum_{i\in\mathcal{N}}\frac{V_i(\mathbf{X})}{V_i(\mathbf{X}^*)}\\
    &=\max_{\mathbf{X}\in \mathcal{X}_f}\sum_{j\in\mathcal{B}}\sum_{c\in\mathcal{C}}\sum_{i\in\mathcal{N}}\frac{R_{ij}^{(c)}x_{ij}^{(c)}}{\sum_{k\in\mathcal{B}}\sum_{l\in\mathcal{C}}R_{ik}^{(l)}{x_{ik}^{(l)}}^{*}}\\
    &=\max_{\mathbf{X}\in \mathcal{X}_f}\sum_{j\in\mathcal{B}}\sum_{c\in\mathcal{C}}x_{ij}^{(c)}\sum_{i\in\mathcal{N}}\frac{R_{ij}^{(c)}}{\sum_{k\in\mathcal{B}}\sum_{l\in\mathcal{C}}R_{ik}^{(l)}{x_{ik}^{(l)}}^{*}}\\
    &=\sum_{j\in\mathcal{B}}\sum_{c\in\mathcal{C}}\max_{\mathbf{X}\in\mathcal{X}_f}x_{ij}^{(c)}\sum_{i\in\mathcal{N}}\frac{R_{ij}^{(c)}}{\sum_{k\in\mathcal{B}}\sum_{l\in\mathcal{C}}R_{ik}^{(l)}{x_{ik}^{(l)}}^{*}}\\
    & = \sum_{j\in\mathcal{B}}\sum_{c\in\mathcal{C}}\max_{i\in\mathcal{N}}\frac{R_{ij}^{(c)}}{\sum_{k\in\mathcal{B}}\sum_{l\in\mathcal{C}}R_{ik}^{(l)}{x_{ik}^{(l)}}^{*}}.
\end{align}
Finally, the contrapositive of the aforementioned proposition gives us `if $N<\sum_{j\in\mathcal{B}}\sum_{c\in\mathcal{C}}\max_{i\in\mathcal{N}}$ $\frac{R_{ij}^{(c)}}{\sum_{k\in\mathcal{B}}\sum_{l\in\mathcal{C}}R_{ik}^{(l)}{x_{ik}^{(l)}}^{*}}$, then
$(\bm{\mu}^*\bm{\nu}^*)$ does not satisfy KKT condition \eqref{eq:KKT_condition_b}'.

According to i) and ii), the solution $(\bm{\mu}^*,\bm{\nu}^*)$ does not satisfies the KKT condition \eqref{eq:KKT_condition_b} if and only if there exists ${j\in\mathcal{B}}$ such that $\nu_j^*>0$, or $\bm{\nu}^*=\bm{0}$ and $N<\sum_{j\in\mathcal{B}}\sum_{c\in\mathcal{C}}\max_{i\in\mathcal{N}}\frac{R_{ij}^{(c)}}{\sum_{k\in\mathcal{B}}\sum_{l\in\mathcal{C}}R_{ik}^{(l)}{x_{ik}^{(l)}}^*}$.
\end{IEEEproof}


Though happening with small probability, if the dual solution can satisfy the KKT conditions, then the solution becomes globally optimal, not just locally optimal, as shown by the following lemma.
\begin{lem}\label{lem:non_diff}
If the solution $(\bm{\mu}^*,\bm{\nu}^*)$ satisfies the KKT condition of the dual problem \eqref{eq:min_g_mu_nu}, then $\mathbf{X}^*$ is the global optimal solution of the problem \eqref{eq:P_UA}.
\end{lem}
\begin{IEEEproof}
Suppose that there exists a global optimal solution of the problem \eqref{eq:P_UA} and $(\boldsymbol{\mu}^*, \boldsymbol{\nu}^*)$ satisfies the KKT conditions  \eqref{eq:KKT_condition_b} of the problem \eqref{eq:min_g_mu_nu}.
Then, $\bm{\nu}^*=\bm{0}$ because $\left.\frac{\partial g(\boldsymbol{\mu}, \boldsymbol{\nu})}{\partial\nu_j}\right|_{\bm{\nu}=\bm{\nu}^*}\neq0$ for all $j\in\mathcal{B}$, as shown in \eqref{eq:dg_dnu}. Suppose that the global maximum point $\mathbf{X}^g$ exists. Because $\bm{\nu}^*=\bm{0}$, we need to have  $N\geq\sum_{i\in\mathcal{N}}\frac{V_i(\mathbf{X}^g)}{V_i(\mathbf{X}^*)}$ as shown in \eqref{eq:condition_N}.
Because the objective function of the problem \eqref{eq:P_UA} is $\log\prod_{i\in\mathcal{N}}V_i(\mathbf{X})$, we have
\begin{equation}
\prod_{i\in\mathcal{N}}\frac{V_i(\mathbf{X}^g)}{V_i(\mathbf{X}^*)}\geq1, \hspace{10pt}\forall \mathbf{X^*}\in\mathcal{X}_f.
\end{equation}
We further have
\begin{equation}\label{eq:condition_T2}
    \frac{\sum_{i\in\mathcal{N}}\frac{V_i(\mathbf{X}^g)}{V_i(\mathbf{X}^*)}}{N}\overset{(a)}\geq\left(\prod_{i\in\mathcal{N}}\frac{V_i(\mathbf{X}^g)}{V_i(\mathbf{X}^*)}\right)^\frac{1}{N}{\geq}1,\hspace{10pt}\forall \mathbf{X^*}\in\mathcal{X}_f,
\end{equation}
where equality (a) holds if only if $\mathbf{X}^*=\mathbf{X}^g$. This gives us
\begin{equation}\label{eq:lem3_result}
    \sum_{i\in\mathcal{N}}\frac{V_i(\mathbf{X}^g)}{V_i(\mathbf{X}^*)}\geq N,\hspace{5pt} \forall \mathbf{X}^*\in\mathcal{X}_f.
\end{equation}
From \eqref{eq:condition_N}, we have $\sum_{i\in\mathcal{N}}\frac{V_i(\mathbf{X}^g)}{V_i(\mathbf{X}^*)}\leq N$, which, combined with \eqref{eq:lem3_result}, yields $\sum_{i\in\mathcal{N}}\frac{V_i(\mathbf{X}^g)}{V_i(\mathbf{X}^*)}= N$. Therefore, we get
\begin{equation}
    \frac{\sum_{i\in\mathcal{N}}\frac{V_i(\mathbf{X}^g)}{V_i(\mathbf{X}^*)}}{N}=\left(\prod_{i\in\mathcal{N}}\frac{V_i(\mathbf{X}^g)}{V_i(\mathbf{X}^*)}\right)^\frac{1}{N}=1,\hspace{10pt}\forall \mathbf{X^*}\in\mathcal{X}_f,
\end{equation}
and hence $\mathbf{X}^*=\mathbf{X}^g$.
Therefore, if $\mathbf{X}^*$ satisfies the KKT conditions \eqref{eq:KKT_condition_b}, it is the global optimal solution $\mathbf{X}^g$ of the problem \eqref{eq:P_UA}.
\end{IEEEproof}

From Lemma \ref{lem:non_diff}, if the solution of the dual problem satisfies the KKT conditions, it gives us the global solution. However, this happens with small probability for varying channels, which can be shown by numerical simulations. Furthermore, not every global maximizer satisfies the condition \eqref{eq:condition_N}, which can be shown by simple counter examples. As a result, an optimal solution in general is difficult to obtain from the dual problem approach with non-exponential computational complexity.
In the next section, we propose an alternative approach to solve the original problem \eqref{eq:P_UA} with feasible-time computational complexity but achieving much higher proportional fairness.

\section{Proposed UA, RA, and PC Algorithm \label{sec:prop}}
\subsection{Proposed UA and RA Algorithm for Given PC}\label{sub:Heuristic}
We first investigate the optimality condition of a mixed-integer maximization problem,
starting with the following definition.
\begin{definition}\cite{Y_Xia_ring_sol}[Sufficient conditions on local optimality]\label{def:mixed_2}
Suppose that $\mathcal{D}$ is a domain of a combinatorial problem. Then, $\mathbf{X}\in\mathcal{D}$ satisfying the following conditions is defined as a $p$-distance ($1\leq p \leq n$) ring solution:
\begin{align}
    q(\mathbf{X}) \geq q(\mathbf{X}^{\prime}), \hspace{10pt}\forall \mathbf{X}^{\prime} \in \left\{ \left\Vert \mathbf{X}-\mathbf{X}^{\prime}\right\Vert _{0}=p, \mathbf{X}^{\prime} \in \mathcal{D} \right\}
\end{align}
where $\left\Vert\mathbf{X}-\mathbf{X^{\prime}}\right\Vert_{0}$ denotes the number of different elements of $\mathbf{X}$ and $\mathbf{X^{\prime}}$.
\end{definition}

The 2-distance ring solution of the problem \eqref{eq:P_UA} without the backhaul constraint \eqref{eq:backhaul_ineq} is derived in Theorem \ref{thm:p-distnace ring solution}.
\begin{theorem}\label{thm:p-distnace ring solution}
The indicator $\mathbf{X}$, which satisfies the conditions \eqref{eq:mu update} and \eqref{eq:x_Allo4},
is a $2$-distnace ring solution of the problem \eqref{eq:P_UA} without the constraint \eqref{eq:backhaul_ineq}. In addition, for any given $\mathbf{X}_{[0]}\in\mathcal{X}_f$, $\mathbf{X}_{[t]}$ in \eqref{eq:x_update}, $t=1, 2, \ldots$, converges to the $2$-distance ring solution $\mathbf{X}^*$.
\begin{equation}\label{eq:mu update}
\zeta_{ij}^{(c)}(\mathbf{X})=\left.\frac{1}{\sum_{j'\in\mathcal{B}}\sum_{c'\in\mathcal{C}}R_{ij'}^{(c')}{x_{ij'}^{(c')}}}\right|_{{x_{ij}^{(c)}}=0},\hspace{5pt}\forall i\in\mathcal{N},\forall j\in\mathcal{B},\forall c\in\mathcal{C},
\end{equation}
\begin{equation}\label{eq:x_Allo4}
{x_{ij}^{(c)}}=\mathbf{1}_{\{i=\argmax_{i'\in\mathcal{N}}R_{i'j}^{(c)}\zeta_{i'j}^{(c)}(\mathbf{X})\}},\hspace{5pt} \forall i\in\mathcal{N},\forall j\in\mathcal{B},\forall c\in\mathcal{C},
\end{equation}
\begin{equation}\label{eq:x_update}
x_{ij,[t]}^{(c)}=\begin{cases}\mathbf{1}_{\{i=\argmax_{i'\in\mathcal{N}}R_{i'j}^{(c)}\zeta_{i'j}^{(c)}(\mathbf{X}_{[t-1]})\}}, &\textrm{if}~j=\left(\left\lfloor\frac{t}{C}\right\rfloor\mod J\right)+1,c=(t\mod C)+1,  \\
x_{ij,[t-1]}^{(c)}, & \textrm{otherwise,}
\end{cases}
\end{equation}
where $\mathbf{X}_{[t]}\in\mathcal{X}_f$ with $[\mathbf{X}_{[t]}]_{ijc}=x_{ij,[t]}^{(c)}$.
\end{theorem}
\begin{IEEEproof}
i)
Note that the proportional fairness in \eqref{eq:prop_fairness} is different from all $\mathbf{X}$ for given $\mathbf{P}$, because $R_{ij}^{(c)}$ is continuously distributed.
A set of $2$-distance points from $\mathbf{X}^*$ is defined as
\begin{equation}\label{eq:2_dist_point}
    A(\mathbf{X}^*)=\{\mathbf{X}|\left\|\mathbf{X}-\mathbf{X}^*\right\|_0=2,\mathbf{X}\in\mathcal{X}_f\}.
\end{equation}
For all $\mathbf{X}\in A(\mathbf{X}^*)$, ${x_{ij}^{(c)}}={x_{ij}^{(c)}}^*$ for all $(i,j,c)\in\mathcal{N}\times\mathcal{B}\times\mathcal{C}$ except for $x_{i'j'}^{(c')} \neq {x_{i'j'}^{(c')}}^*$ and $x_{i''j'}^{(c')} \neq {x_{i''j'}^{(c')}}^*$ for some $(j^\prime,c^\prime)$ and $i''\neq i'$.
Let $\mathbf{X}^*$ satisfies the conditions \eqref{eq:x_Allo4}. Then, we have
\begin{equation}\label{eq:ineq_zeta}
    \sum_{i\in\mathcal{N}}R_{ij^\prime}^{(c^\prime)}\zeta_{ij}^{(c^\prime)}(\mathbf{X}^*){x_{ij^\prime}^{(c^\prime)}}^*> \sum_{i\in\mathcal{N}}R_{ij^\prime}^{(c^\prime)}\zeta_{ij^\prime}^{(c^\prime)}(\mathbf{X^*})x_{ij^\prime}^{(c^\prime)}, \hspace{5pt}\forall \mathbf{X}\in A(\mathbf{X}^*),
\end{equation}
where $\zeta_{ij}^{(c)}$ is represented in \eqref{eq:mu update}.
A function $\omega_{j}^{(c)}(\mathbf{X})$ is defined as
\begin{equation}
    \omega_{j}^{(c)}(\mathbf{X}) = \left.\prod_{i\in\mathcal{N}}\left(\sum_{m\in\mathcal{B}}\sum_{l\in\mathcal{C}}R_{im}^{(l)}x_{im}^{(l)}\right)\right|_{x_{ij}^{(c)}=0,\forall i\in\mathcal{N}}.
\end{equation}
Then, the proportional fairness function $U(\mathbf{X},\mathbf{P})$ in \eqref{eq:prop_fairness} with given $\mathbf{P}$ is represented as
\begin{equation} \label{eq:UX}
    e^{U(\mathbf{X},\mathbf{P})} = \omega_{j}^{(c)}(\mathbf{X})\left(1+\sum_{i\in\mathcal{N}}R_{ij}^{(c)}\zeta_{ij}^{(c)}(\mathbf{X})x_{ij}^{(c)}\right),\hspace{5pt}\forall j\in\mathcal{B}, \forall c\in\mathcal{C}.
\end{equation}
\pagebreak[0]From \eqref{eq:UX}, we have
\pagebreak[0]
\begin{align}
    \displaybreak[0]\frac{e^{U(\mathbf{X}^*,\mathbf{P})}}{e^{U(\mathbf{X},\mathbf{P})}} =&\frac{\omega_{j^\prime}^{(c^\prime)}(\mathbf{X}^*)\left(1+\sum_{i\in\mathcal{N}}R_{ij'}^{(c')}\zeta_{ij^\prime}^{(c^\prime)}(\mathbf{X}^*){x_{ij^\prime}^{(c^\prime)}}^*\right)}{\omega_{j^\prime}^{(c^\prime)}(\mathbf{X})\left(1+\sum_{i\in\mathcal{N}}R_{ij'}^{(c')}\zeta_{ij^\prime}^{(c^\prime)}(\mathbf{X})x_{ij'}^{(c')}\right)}\\
    \displaybreak[0]=&\frac{\omega_{j^\prime}^{(c^\prime)}(\mathbf{X}^*)\left(1+\sum_{i\in\mathcal{N}}R_{ij'}^{(c')}\zeta_{ij^\prime}^{(c^\prime)}(\mathbf{X}^*){x_{ij^\prime}^{(c^\prime)}}^*\right)}{\omega_{j^\prime}^{(c^\prime)}(\mathbf{X}^*)\left(1+\sum_{i\in\mathcal{N}}R_{ij'}^{(c')}\zeta_{ij^\prime}^{(c^\prime)}(\mathbf{X}^*)x_{ij'}^{(c')}\right)}> 1,\label{eq:UX_ratio2} \hspace{5pt}\forall \mathbf{X}\in A(\mathbf{X^*}),
\end{align}
which follows from the fact that $\mathbf{X} = \mathbf{X}^*$ except for $x_{i'j'}^{(c')} \neq {x_{i'j'}^{(c')}}^*$ and $x_{i''j'}^{(c')} \neq {x_{i''j'}^{(c')}}^*$ for some $(j^\prime,c^\prime)$ and $i''\neq i'$, and by the definition of $U(\mathbf{X},\mathbf{P})$ in \eqref{eq:UX}.
From \eqref{eq:ineq_zeta} and \eqref{eq:UX_ratio2},
\begin{equation}
    U(\mathbf{X}^*,\mathbf{P})>U(\mathbf{X},\mathbf{P}), ~\forall \mathbf{X}\in A(\mathbf{X}^*).
\end{equation}
Therefore, $\mathbf{X}^*$ is a $2$-distance ring solution if $\mathbf{X}^*$ satisfies \eqref{eq:mu update}-\eqref{eq:x_Allo4}.
\\
ii)
Note that $\mathbf{X}_{[t]}\in A(\mathbf{X}_{[t-1]]})\cup\{\mathbf{X}_{[t-1]}\}$.
The definition of $\mathbf{X}_{[t]}$ in \eqref{eq:x_update} gives us
\begin{equation}\label{eq:ineq_U}
\sum_{i\in\mathcal{N}}R_{i\tilde{j}}^{(\tilde{c})}\zeta_{i\tilde{j}}^{(\tilde{c})}(\mathbf{X}_{[t-1]})x_{i\tilde{j},[t]}^{(\tilde{c})}\geq\sum_{i\in\mathcal{N}}R_{i\tilde{j}}^{(\tilde{c})}\zeta_{i\tilde{j}}^{(\tilde{c})}(\mathbf{X}_{[t-1]})x_{i\tilde{j},[t-1]}^{(\tilde{c})},~\forall t\ge 1,
\end{equation}
where $\tilde{j}=\left(\left\lfloor\frac{t}{C}\right\rfloor\mod J\right)+1$ and $\tilde{c}=(t\mod C)+1$. Then, from \eqref{eq:UX} and \eqref{eq:ineq_U}, we have
\begin{equation}\label{eq:U_frac}
\frac{e^{U(\mathbf{X}_{[t]},\mathbf{P})}}{e^{U(\mathbf{X}_{[t-1]},\mathbf{P})}}=\frac{\omega_{\tilde{j}}^{(\tilde{c})}(\mathbf{X}_{[t]})\left(1+\sum_{i\in\mathcal{N}}R_{i\tilde{j}}^{(\tilde{c})}\zeta_{i\tilde{j}}^{(\tilde{c})}(\mathbf{X}_{[t-1]})x_{i\tilde{j},[t]}^{(\tilde{c})}\right)}{\omega_{\tilde{j}}^{(\tilde{c})}(\mathbf{X}_{[t-1]})\left(1+\sum_{i\in\mathcal{N}}R_{i\tilde{j}}^{(\tilde{c})}\zeta_{i\tilde{j}}^{(\tilde{c})}(\mathbf{X}_{[t-1]})x_{i\tilde{j},[t-1]}^{(\tilde{c})}\right)}\geq1,\hspace{5pt}\forall t\ge 1.
\end{equation}
Let $\mathbf{Y}_n$ be defined as $\mathbf{Y}_n=\mathbf{X}_{[nJC]}$ for all $n=0, 1, \ldots$. Then, from \eqref{eq:U_frac},  $U(\mathbf{Y}_n,\mathbf{P})$ is monotonically increasing with respect to $n$, i.e., $U(\mathbf{Y}_n,\mathbf{P})\leq U(\mathbf{Y}_{n+1},\mathbf{P})$ for all $n=0,1,\ldots$.
Because the resultant proportional fairness is different for all $\mathbf{X}\in\mathcal{X}_f$,
 $U(\mathbf{Y}_n,\mathbf{P})$ is strictly increasing for all $n\le k$, where $k =\argmin_{m}\{m|\mathbf{Y}_m=\mathbf{Y}_{m+1}\}$. From the result in the part i),  $\mathbf{Y}_n$ is a $2$-distance ring solution of the problem if $U(\mathbf{Y}_{n},\mathbf{P})=U(\mathbf{Y}_{n+1},\mathbf{P})$. Therefore, $\mathbf{Y}_{n}$ converges to a local optimal 2-distance ring solution, which shows that $\mathbf{X}_{[t]}$ converges to a 2-distance ring solution $\mathbf{X}^*$.
\end{IEEEproof}
From Theorem \ref{thm:p-distnace ring solution}, the local optimal 2-distance ring solution $\mathbf{X}^*$ is obtained from \eqref{eq:x_update} for the problem \eqref{eq:P_UA} without the consideration of the backhaul constraint. Now, the aim is to take into account the backhaul constraint. In the dual problem approach, the lagrange variable $\nu_{j}$ for the $j$-th backhaul constraint plays a role of pricing in \eqref{eq:dual function} on the amount of data transmission of SBS $j$. Specifically, if the amount of backhaul left at SBS $j$  becomes small, $\nu_{j}$ increases, resulting in negative $R_{ij}^{(c)}(\mu_i-\nu_j)$ in the cost function of \eqref{eq:dual function}. Thus, the user $i$ with large $R_{ij}^{(c)}$, $c\in \mathcal{C}$, shall have larger magnitude of $R_{ij}^{(c)}(\mu_i-\nu_j)$ with a negative sign, which results in less chance to be allocated for RB $c$ of SBS $j$ in the max operation of \eqref{eq:dual function}. Therefore, users with smaller rates $R_{ij}^{(c)}$ are allocated for SBS $j$ so that the backhaul constraint \eqref{eq:backhaul_ineq} of SBS $j$ is satisfied.

\pagebreak[0]Inspired by the pricing approach, we modify the conditions \eqref{eq:mu update} and \eqref{eq:x_Allo4} as follows:
\begin{equation} \label{eq:zeta2}
\displaybreak[0]\zeta_{ij}^{(c)}(\mathbf{X}^*)=\left.\frac{1}{\sum_{j'\in\mathcal{B}}\sum_{c'\in\mathcal{C}}R_{ij'}^{(c')}{x_{ij'}^{(c')}}^*}\right|_{{x_{ij}^{(c)}}^*=0}, \hspace{5pt}\forall i\in\mathcal{N},\forall j\in\mathcal{B},\forall c\in\mathcal{C},
\end{equation}
\begin{equation} \label{eq:x_Alloc5}
\displaybreak[0]{x_{ij}^{(c)}}^*=\mathbf{1}_{\{i=\argmax_{i'\in\mathcal{N}}R_{i'j}^{(c)}(\zeta_{i'j}^{(c)}(\mathbf{X}^*)-\nu_{j})\}} ,\hspace{5pt}\forall i\in\mathcal{N},\forall j\in\mathcal{B},\forall c\in\mathcal{C}.
\end{equation}
The lagrange multiplier $\nu_j$ is imposed for pricing the data rate of SBS $j$ and thereby satisfying the backhaul constraint. Specifically, $\nu_j$ is obtained from the sub-gradient method as
\begin{align}\label{eq:nu_update_2}
\nu_{j}&\coloneqq\left[\nu_{j}-\alpha\frac{\partial g(\boldsymbol{\mu}, \boldsymbol{\nu})}{\partial \nu_j}\right]^{+}=\left[\nu_{j}-\alpha\left(Z_{j}-\sum_{i\in\mathcal{N}}\sum_{c\in\mathcal{C}}R_{ij}^{(c)}{x_{ij}^{(c)}}^*\right)\right]^{+},
\end{align}
where $\alpha$ denotes the step size.

An alternative of sub-gradient method \eqref{eq:nu_update_2} is also proposed based on the cyclic coordinate descent method for faster convergence. In the cyclic coordinate descent method, only one variable from $\boldsymbol{\nu}_j$ is sequentially updated with the other variables fixed. That is, $\nu_j$ at the $(t+1)$-th iteration, denoted by $\nu_j^{(t+1)}$, is updated by
\begin{equation}
\nu_{j}^{(t+1)}=\argmin_{\gamma\geq0}g(\nu_{1}^{(t)},\ldots,\nu_{j-1}^{(t)},\gamma,\nu_{j+1}^{(t)},\ldots,\nu_{J}^{(t)}),\label{eq:nu_update}
\end{equation}
where $\gamma$ is obtained by sub-gradient method for $\nu_{j}$ in \eqref{eq:nu_update_2}.
The details of the cyclic coordinate descent algorithm is presented
in Algorithm \ref{alg:Cyclic coordinate descent}.

Note that Algorithm \ref{alg:Cyclic coordinate descent} finds the 2-distance ring solution, because the conditions \eqref{eq:zeta2} and \eqref{eq:x_Alloc5} are identical to \eqref{eq:mu update} to \eqref{eq:x_Allo4} if $\nu_j=0$, i.e., the backhaul constraint is strictly satisfied. If the backhaul constraint of SBS $j$ is not satisfied for previously found $\mathbf{X}$, $\nu_j$ is updated by a positive value of $-\alpha\left(Z_{j}-\sum_{i\in\mathcal{N}}\sum_{c\in\mathcal{C}}R_{ij}^{(c)}x_{ij}^{(c)*}\right)$. At the next iteration then, $(\zeta_{ij}^{(c)}(\mathbf{X}^*)-\nu_{j})$ in \eqref{eq:x_Alloc5} may become negative. As a result, users with smaller rates $R_{ij}^{(c)}$ are selected for SBS $j$ so that the backhaul constraint is satisfied.

In fact, unlike in Theorem \ref{thm:p-distnace ring solution}, the sub-gradient or cyclic coordinate descent method does not always guarantee a 2-distance ring solution due to the additional backhaul constraint. Hence, a gap between the solution of Algorithm \ref{alg:Cyclic coordinate descent} and 2-distance points of the solution is derived in Lemma \ref{lem:gap_bound}.
 \begin{lem}\label{lem:gap_bound}
 The gap of the proportional fairness between the solution of Algorithm \ref{alg:Cyclic coordinate descent} $\mathbf{X}^*$ and 2-distance points from the solution is bounded by  $\max_{j\in\mathcal{B}}\epsilon_{j}\nu_{j}$, where $\epsilon_{j} = Z_{j}- \sum_{i\in\mathcal{N}}\sum_{c\in\mathcal{C}}R_{ij}^{(c)}{x_{ij}^{(c)}}^*$. which is $0$ if $C\rightarrow\infty$.
 \end{lem}
 \begin{IEEEproof}
 The proof is shown in Appendix \ref{sec:Appendix_gap}.
 \end{IEEEproof}
From Lemma \ref{lem:gap_bound}, as the number of RBs per SBS increases, the local optimal solution is asymptotically guaranteed even with the modified optimality conditions \eqref{eq:zeta2} and \eqref{eq:x_Alloc5}, i.e., Algorithm \ref{alg:Cyclic coordinate descent}, where $x_{ij}^{(c)}$ and $\zeta_{ij}^{(c)}$ are sequentially updated. In this sequential update,
because $\mathbf{X}$ does not have any changes if $\nu_j$ changed with very small amount, the closed-form update for $\nu_j$ is adjusted.
Let $\mathbf{X}_{\textrm{old}}\in\mathcal{X}_f$ with $[\mathbf{X}_{\textrm{old}}]_{ijc}=x_{ij,\textrm{old}}^{(c)}$ satisfy the conditions \eqref{eq:zeta2} and \eqref{eq:x_Alloc5}. Then, resource allocation of SBS $j$ on RB $c$ is changed into user $i'$, i.e., $x_{ij}^{(c)}:1\rightarrow0$ and $x_{i'j}^{(c)}:0\rightarrow1$, if $R_{i'j}^{(c)}(\zeta_{i'j}^{(c)}(\mathbf{X}_{\textrm{old}})-\nu_j)\ge\sum_{i\in\mathcal{N}}R_{ij}^{(c)}(\zeta_{ij}^{(c)}(\mathbf{X}_{\textrm{old}})-\nu_j)x_{ij,\textrm{old}}^{(c)}$. Then, the $\nu_j$ value which changes allocation of SBS $j$ on RB $c$ into user $i'$ is denoted
\begin{equation}
\beta_{i',j,c}=\frac{R_{i'j}^{(c)}\zeta_{i'j}^{(c)}(\mathbf{X}_{\textrm{old}})-\sum_{i\in\mathcal{N}}R_{ij}^{(c)}\zeta_{ij}^{(c)}(\mathbf{X}_{\textrm{old}})x_{ij,\textrm{old}}^{(c)}}{R_{i'j}^{(c)}-\sum_{i\in\mathcal{N}}R_{ij}^{(c)}x_{ij,\textrm{old}}^{(c)}},\hspace{5pt}\forall i\in\mathcal{N},\forall j\in\mathcal{B},\forall c\in\mathcal{C}.\label{eq:Beta}\end{equation}
For any $j\in\mathcal{B}$, because the sign of gradient step for $\nu_j$ is same with the sign of $\sum_{i\in\mathcal{N}}\sum_{c\in\mathcal{C}}R_{ij}^{(c)}x_{ij}^{(c)}-Z_j$, the nearest $\beta_{ij}^{(c)}$ from $\nu_j$ is denoted as
\begin{equation}\label{eq:dynamic}
\nu_{j}^{(\textrm{new})}=\begin{cases}
\min_{i,c}\left.\beta_{i,j,c}\right|_{\beta_{i,j,c}>\nu_{j}}, & \text{if}\,Z_{j}\leq\sum_{i\in\mathcal{N}}\sum_{c\in\mathcal{C}}R_{ij}^{(c)}x_{ij}^{(c)},\\
\max_{i,c}\left.\beta_{i,j,c}\right|_{\beta_{i,j,c}<\nu_{j}}, & \text{otherwise}.
\end{cases}
\end{equation}
Then, because the aim is to find a nearest value of $\nu_j$ that changes $\mathbf{X}$, In other words, the step size $\alpha$ in \eqref{eq:nu_update_2} is replaced by dynamic step size $\alpha_{\textrm{dynamic}}$ as follow:
\begin{equation}
   \alpha_{\textrm{dynamic}} = \max\left(\alpha,\left|\frac{\nu_j-\nu_j^{(\textrm{new})}}{Z_j-\sum_{i\in\mathcal{N}}\sum_{j\in\mathcal{B}}R_{ij}^{(c)}x_{ij}^{(c)}}\right|\right).
\end{equation}

\begin{algorithm}
\caption{Cyclic coordinate descent method for the proposed UA and RA algorithm\label{alg:Cyclic coordinate descent}}

\textbf{~~~Initialization : }set $\nu_{j}=0$, $\forall j$. set
$\zeta_{ij}^{(c)}=0,\forall i$.

~~~\textbf{repeat }

~~~~~~\textbf{for $\forall j\in\mathcal{B}$}

~~~~~~~~~\textbf{repeat }

~~~~~~~~~~~~\textbf{for $\forall c\in C$}

~~~~~~~~~~~~~~~1) Update $x_{ij}^{(c)}$ according
to \eqref{eq:x_Alloc5}.

~~~~~~~~~~~~~~~2) Update $\zeta_{ij}^{(c)}$ according to \eqref{eq:zeta2}.

~~~~~~~~~~~~\textbf{end}

~~~~~~~~~\textbf{until $\zeta_{ij}^{(c)}$ }converges

~~~~~~~~~3) Update $\nu_{j}$ according to \eqref{eq:nu_update}.

~~~~~~\textbf{end}

~~~\textbf{until } $\nu_{j}$ converges and $Z_j\geq\sum_{i\in\mathcal{N}}\sum_{c\in\mathcal{C}}R_{ij}^{(c)}x_{ij}^{(c)}, \forall j\in\mathcal{B}$.

~~~\textbf{Return : }indicator \textbf{$x_{ij}^{(c)}$}
\end{algorithm}
\subsection{Proposed Power Control Algorithm for Given UA and RA\label{sub:Power-allocation}}
In Section \ref{sub:Heuristic}, the RA and UA for maximizing the proportional fairness is considered for given $\mathbf{P}$. Here, a per-RB PC is proposed to maximize the proportional fairness by allocating power on each RB with given $\mathbf{X}$.
In Section \ref{sec:dual-problem} and \ref{sub:Heuristic}, the constraint \eqref{eq:sum=00003D1} assumes that each RB is always allocated to one of the users.
However, allocating an RB to no user should also be considered. Fortunately, this can be taken into account by allocating zero power on the RB. Then, the PC problem for given $\mathbf{X}$ is formulated from \eqref{eq:P1} as:
\begin{subequations}\label{eq:problem_PA}
\begin{align}\mathrm{P3: } \max_{\mathbf{P}} & \sum_{i\in\mathcal{N}}\log\sum_{j\in\mathcal{B}}\sum_{c\in\mathcal{C}}R_{ij}^{(c)}x_{ij}^{(c)}\\
\textrm{s.t}\hspace{5pt} & \sum_{c\in\mathcal{C}}\sum_{i\in\mathcal{N}}R_{ij}^{(c)}x_{ij}^{(c)}\leq Z_{j},\hspace{5pt}\forall j\in\mathcal{B}\label{eq:P3_c}\\
 & \sum_{c\in\mathcal{C}}p_{j}^{(c)}\leq P_{j,\textrm{max}},\hspace{5pt}\forall j\in\mathcal{B}\label{eq:P3_d}\\
 & R_{ij}^{(c)}=W\log_{2}\left(1+\frac{\left|h_{ij}^{(c)}\right|^{2}p_{j}^{(c)}}{\sigma^{2}+\sum_{k\neq j}\left|h_{ik}^{(c)}\right|^{2}p_{k}^{(c)}}\right),\hspace{5pt}\forall i\in\mathcal{N},\forall j \in\mathcal{B},\forall c\in\mathcal{C}\\
 & p_{j}^{(c)}\geq0,\hspace{5pt}\forall j \in\mathcal{B},\forall c\in\mathcal{C}\label{eq:P3_f}
\end{align}
\end{subequations}
To solve the problem \eqref{eq:problem_PA}, we first relax the problem without the constraint \eqref{eq:P3_c}, which shall be considered later, as follow:
\begin{subequations}\label{eq:Problem_PA_2}
\begin{align}\mathrm{P4: }\max_{\mathbf{P}} & \sum_{i\in\mathcal{N}}\log\sum_{j\in\mathcal{B}}\sum_{c\in\mathcal{C}}W\log_{2}\left(1+\frac{\left|h_{ij}^{(c)}\right|^{2}p_{j}^{(c)}x_{ij}^{(c)}}{\sigma^{2}+\sum_{k\neq j}\left|h_{ik}^{(c)}\right|^{2}p_{k}^{(c)}}\right)\\
\textrm{s.t}\,\,\, & \sum_{c\in\mathcal{C}}p_{j}^{(c)}\leq P_{j,\textrm{max}},\hspace{5pt}\forall j\in\mathcal{B}\label{eq:P4_b}\\
 & p_{j}^{(c)}\geq0,\hspace{5pt}\forall j \in\mathcal{B},\forall c\in\mathcal{C}.\label{eq:P4_c}
\end{align}
\label{eq:problem_power_simple}
\end{subequations}
The aim is to obtain the KKT conditions of the problem \eqref{eq:problem_power_simple}.
The Lagrangian of the problem \eqref{eq:problem_power_simple} is represented as
\begin{align}\label{eq:L_p}
L(\mathbf{P},\bm{\xi},\bm{\varphi}) & =\sum_{i\in\mathcal{N}}\log\sum_{j\in\mathcal{B}}\sum_{c\in\mathcal{C}}W\log_{2}\left(1+\frac{\left|h_{ij}^{(c)}\right|^{2}p_{j}^{(c)}x_{ij}^{(c)}}{\sigma^{2}+\sum_{k\neq j}\left|h_{ik}^{(c)}\right|^{2}p_{k}^{(c)}}\right)\\
 & +\sum_{j\in\mathcal{B}}\xi_{j}\left(P_{j,\textrm{max}}-\sum_{c\in\mathcal{C}}p_{j}^{(c)}\right)+\sum_{j\in\mathcal{B}}\sum_{c\in\mathcal{C}}\varphi_{j}^{(c)}p_{j}^{(c)},
\end{align}
where $\bm{\xi}\in\mathbb{R}_{\ge0}^{B}$ with ${[\bm{\xi}]}_{j}=\xi_{j}$, and $\bm{\varphi}\in\mathbb{R}_{\ge0}^{B\times C}$ with  ${[\bm{\varphi}]}_{jc}=\varphi_{j}^{(c)}$ are lagrange multipliers corresponding to the constraints \eqref{eq:P4_b} and \eqref{eq:P4_c}.
The KKT conditions of the problem \eqref{eq:problem_power_simple} are established in Lemma \ref{lem:KKT_power}.
\begin{lem}\label{lem:KKT_power}
The KKT conditions of the problem \eqref{eq:problem_power_simple} are given by
\begin{subnumcases}{\label{eq:KKT2}}
\frac{\partial U(\mathbf{X},\mathbf{P})}{\partial p_{j}^{(c)}} = \xi_j,\hspace{5pt} \forall (j,c)\in\mathcal{B}\times\mathcal{C} \backslash{\{(k,l)|p_{k}^{(l)}=0\}} \label{eq:KKT2_a} \\
\frac{\partial U(\mathbf{X},\mathbf{P})}{\partial p_{j}^{(c)}} \leq \xi_j,\hspace{5pt} \forall j\in\mathcal{B},\forall c\in\mathcal{C} \label{eq:KKT2_b}\\
p_{j}^{(c)} \geq 0,\hspace{5pt} \forall j\in\mathcal{B}, c\in\mathcal{C}\\
P_{j,\textrm{max}}-\sum_{c\in\mathcal{C}}p_{j}^{(c)}\geq 0,\hspace{5pt} \forall j\in\mathcal{B}\\
\xi_j\geq 0, \hspace{5pt}\forall j\in\mathcal{B}\label{eq:KKT2_e} \\
\xi_j(P_{j,\textrm{max}}-\sum_{c\in\mathcal{C}}p_{j}^{(c)}) = 0,\hspace{5pt} \forall j\in\mathcal{B} \label{eq:KKT2_f}
\end{subnumcases}
\end{lem}
\begin{IEEEproof}
Proof in Appendix \ref{sec:Appendix_B}
\end{IEEEproof}

Now, we propose an algorithm to find a solution that satisfies the KKT conditions in Lemma \ref{lem:KKT_power} with the consideration of the backhaul constraint  \eqref{eq:P3_c}. We first start with the following theorem.

\begin{lem}\label{lem:local_opt}
For a local optimal point of the problem \eqref{eq:Problem_PA_2}, there exists at least one  $j\in\mathcal{B}$ which satisfies  $\sum_{c\in\mathcal{C}}p_{j}^{(c)}=P_{j,\textrm{max}}$.
\end{lem}
\begin{IEEEproof}
We only consider $\mathbf{P}\neq \boldsymbol{0}$, because $\mathbf{P}= \boldsymbol{0}$ cannot be a local optimal solution.
For convenience, let us vectorize $\mathbf{P}\in \mathbb{R}^{B\times C}$ as $\mathbf{p}=\textrm{vect}(\mathbf{P}) \in \mathbb{R}^{B\cdot C \times 1}$.
For any $\mathbf{w} \in \mathbb{C}^{B\cdot C \times 1}$ with $\Vert\mathbf{w}\Vert_2=1$, the directional derivative of the function $U(\mathbf{X},\mathbf{p})$ is represented as
\begin{equation}
D_{\mathbf{w}}(U) = \lim_{\Delta\rightarrow0^{+}}\frac{U(\mathbf{X},\mathbf{p}+\Delta\cdot \mathbf{w})-U(\mathbf{p})}{\Delta} = \mathbf{w}^{\mathrm{T}}\nabla_{\mathbf{p}} U(\mathbf{X},\mathbf{p}),
\end{equation}
where $\nabla_{\mathbf{p}} U(\mathbf{X},\mathbf{p})=\left[ \frac{\partial U(\mathbf{X}, \mathbf{p})}{\partial p_{1}^{(1)}},  \frac{\partial U(\mathbf{X}, \mathbf{p})}{\partial p_{1}^{(2)}}, \ldots, \frac{\partial U(\mathbf{X}, \mathbf{p})}{\partial p_{B}^{(C)}} \right]^{\mathrm{T}}$.
Let $\mathcal{P}_f$ denote the set of $\mathbf{p}$'s satisfying the constraints \eqref{eq:P4_b} and \eqref{eq:P4_c}.
Then, for all $\mathbf{\tilde{p}}\in\mathcal{P}_f$ except for the trivial case $\mathbf{\tilde{p}}=\boldsymbol{0}$, choosing the direction as $\mathbf{w}=\frac{\mathbf{\tilde{p}}}{{\Vert\mathbf{\tilde{p}}\Vert}_2}$ gives us
\begin{align}
D_{\mathbf{w}}(U)&=\frac{1}{{\Vert \mathbf{\tilde{p}}\Vert}_{2}}\mathbf{\tilde{p}}^{\mathrm{T}}\nabla_{\mathbf{p}} U(\mathbf{X},\mathbf{\tilde{p}})\label{eq:direct_deriv2}\\
&=\lim_{\Delta\rightarrow0^{+}}\frac{U\left(\mathbf{X},\mathbf{\tilde{p}}+\Delta\cdot\frac{\mathbf{\tilde{p}}}{{\Vert\mathbf{\tilde{p}}\Vert}_2}\right)-U(\mathbf{\tilde{p}})}{\Delta}\label{eq:direct_deriv3}\\
&=\left.\frac{1}{{\Vert \mathbf{\tilde{p}}\Vert}_{2}}\cdot \frac{\partial U(\mathbf{X},\mathbf{\tilde{p}}(1+\Delta))}{\partial \Delta}\right|_{\Delta = 0},\label{eq:direct_deriv4}
\end{align}
where \eqref{eq:direct_deriv4} follows from the l'H\^{o}pital's law. Then, we further have
\begin{align}
    \left.\frac{\partial U(\mathbf{X},\mathbf{\tilde{p}}(1+\Delta))}{\partial \Delta}\right|_{\Delta = 0}
    =\left.\sum_{i\in\mathcal{N}}\frac{1}{\sum_{k\in\mathcal{B}}\sum_{l\in\mathcal{C}}R_{ik}^{(l)}x_{ik}^{(l)}}\sum_{j\in\mathcal{B}}\sum_{c\in\mathcal{C}}\frac{\partial R_{ij}^{(c)}\Big|_{p_{j}^{(c)}=\tilde{p}_{j}^{(c)}(1+\Delta)}}{\partial \Delta}x_{ij}^{(c)}\right|_{\Delta=0}.\label{eq:partial_Delta1}
\end{align}
Here, we get
\begin{align}
\left.\frac{\partial R_{ij}^{(c)}|_{p_{j}^{(c)}=\tilde{p}_{j}^{(c)}(1+\Delta)}}{\partial \Delta}\right|_{\Delta=0}=\frac{{|h_{ij}^{(c)}|}^2\tilde{p}_{j}^{(c)}}{{(1+\sum_{b\neq j}{|h_{ib}^{(c)}|}^2\tilde{p}_{b}^{(c)}})({1+\sum_{b\in\mathcal{B}}{|h_{ib}^{(c)}|}^2\tilde{p}_{b}^{(c)}})}\ge 0.\label{eq:partial_Delta2}
\end{align}
From \eqref{eq:direct_deriv2} to \eqref{eq:partial_Delta2} and the fact that $\tilde{p}_j^{(c)}\geq 0$ and $\mathbf{p}\neq \mathbf{0}$, we have
\begin{equation}\label{eq:lim_direction}
\lim_{\Delta\rightarrow0^{+}}\frac{U\left(\mathbf{X},\mathbf{\tilde{p}}+\Delta\cdot\frac{\mathbf{\tilde{p}}}{{\Vert\mathbf{\tilde{p}}\Vert}_2}\right)-U(\mathbf{\tilde{p}})}{\Delta}=\frac{1}{{\Vert\mathbf{\tilde{p}}\Vert}_2}\mathbf{\tilde{p}}^{\mathrm{T}}\nabla_{\mathbf{p}} U(\mathbf{X},\mathbf{\tilde{p}})>0.
\end{equation}
From \eqref{eq:lim_direction}, we have
\begin{equation}
    \frac{1}{{\Vert\mathbf{p}\Vert}_2}\mathbf{p}^{\mathrm{T}}\nabla_{\mathbf{p}} U(\mathbf{X},\mathbf{p}) = \frac{1}{{\Vert\mathbf{p}\Vert}_2}\sum_{j\in\mathcal{B}}\sum_{c\in\mathcal{C}} p_{j}^{(c)}\frac{\partial U(\mathbf{X},\mathbf{p})}{\partial p_{j}^{(c)}} >0, \hspace{5pt}\forall \mathbf{p}\in\mathcal{P}_f\backslash\{\mathbf{0}\}.\label{eq:Dw_positive}
\end{equation}
From \eqref{eq:Dw_positive}, since $p_{j}^{(c)}\geq 0$ for all $j\in \mathcal{B}$ and $c\in \mathcal{C}$, there must exist at least one index $(j^*,c^*)$ for some $j^*\in \mathcal{B}$ and $c^*\in \mathcal{C}$ such that \begin{equation}\label{eq:power_positive}
    \frac{\partial U(\mathbf{X},\mathbf{P})}{\partial p_{j^*}^{(c^*)}}>0.
\end{equation}
Therefore, for any $\mathbf{p}\in\mathcal{P}_f$, there exists at least one positive element of $\nabla_{\mathbf{p}} U(\mathbf{X},\mathbf{p})$.

Now, we prove the converse. Let us denote the local optimal solution of the problem \eqref{eq:Problem_PA_2} by $\mathbf{P}^*$. Now, suppose that for all $j\in\mathcal{B}$, $P_{j,\textrm{max}}-\sum_{c\in\mathcal{C}}{p_{j}^{(c)}}^*>0$. Then, since the local optimal solution satisfies the KKT conditions in \eqref{eq:KKT2}, $\xi_{j}=0$ from \eqref{eq:KKT2_f}. In addition, from \eqref{eq:KKT2_b}, we have
\begin{equation}\label{eq:nabla_U}
    \left.\frac{\partial U(\mathbf{X},\mathbf{P})}{\partial p_{j}^{(c)}}\right|_{\mathbf{P}=\mathbf{P}^*} \leq 0, \hspace{5pt}\forall j\in\mathcal{B},\forall c\in\mathcal{C}.
\end{equation}
Because \eqref{eq:nabla_U} contradicts to \eqref{eq:power_positive}, for local optimal solution $\mathbf{P}^*$, it is not possible for all $j\in\mathcal{B}$, $P_{j,\textrm{max}}-\sum_{c\in\mathcal{C}}{p_{j}^{(c)}}^*>0$.
\end{IEEEproof}


\begin{figure}
\includegraphics[width=1\columnwidth]{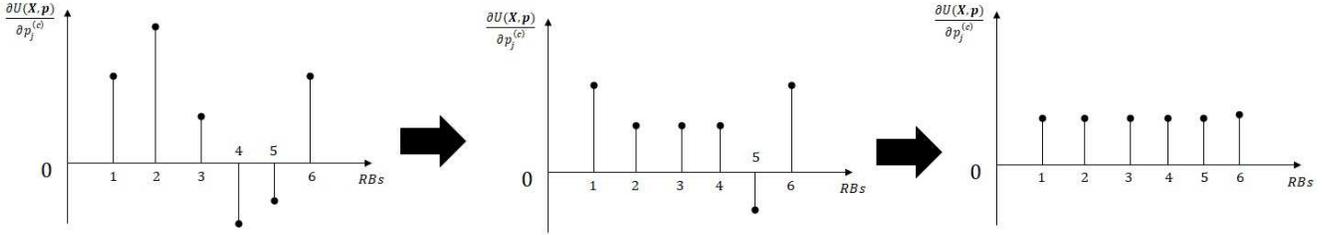}\centering\caption{partial derivatives of the proportional fairness function by each of RBs\label{fig:graph_partial_derivatives}}
\end{figure}
From Lemma \ref{lem:local_opt}, at least one $j\in\mathcal{B}$ exists that satisfies $P_{j,\textrm{max}}=\sum_{c\in\mathcal{C}}p_{j}^{(c)}$.
Let $\mathcal{B}_{\textrm{eq}} = \{j|P_{j,\textrm{max}}=\sum_{c\in\mathcal{C}}p_{j}^{(c)},j\in\mathcal{B}\}$ and $\mathcal{B}_{\textrm{neq}}=\{j|P_{j,\textrm{max}}>\sum_{c\in\mathcal{C}}p_{j}^{(c)},j\in\mathcal{B}\}$. Then, from \eqref{eq:KKT2_a} and \eqref{eq:KKT2_f}, we have $\frac{\partial U(\mathbf{X},\mathbf{P})}{\partial p_{j}^{(c)}} = \xi_j,$ for all $j\in\mathcal{B}_\mathrm{eq},c\in\mathcal{C} \backslash{\{(k,l)|p_{k}^{(l)}=0\}}$ at the optimal point, where $\xi_j>0$. On the other hand, we have $\frac{\partial U(\mathbf{X},\mathbf{P})}{\partial p_{j}^{(c)}} = \xi_{j},$ for all $j\in\mathcal{B}_{\textrm{neq}},c\in\mathcal{C}\backslash{\{(k,l)|p_{k}^{(l)}=0\}}$ at the optimal point, where $\xi_{j}=0$. Therefore, we aim to design $\mathbf{P}$ such that $\frac{\partial U(\mathbf{X},\mathbf{P})}{\partial p_{j}^{(c)}}$ for the RBs with non-zero transmission power become all identical within the same SBS. Since it is difficult to obtain the solution $\mathbf{P}$ in a closed-form, we propose a sequential update of $\mathbf{P}$ from a SBS to another SBS.

Fig. \ref{fig:graph_partial_derivatives} depicts the example of the proposed algorithm in case of SBS $j\in\mathcal{B}_{\textrm{eq}}$, i.e., $\frac{\partial U(\mathbf{X},\mathbf{P})}{\partial p_{j}^{(c)}}$ for all RBs of SBS $j$ with non-zero transmission power are identical to a positive constant. The left figure of Fig. \ref{fig:graph_partial_derivatives} depicts $\frac{\partial U(\mathbf{X},\mathbf{P})}{\partial p_{j}^{(c)}}$ with the initial $\mathbf{P}$. We select two RBs, RB 2 and 4, and take $\Delta p$ from the transmission power for RB 4 to give it to RB 2. For small $\Delta p$, this only makes the cost function $U(\mathbf{X},\mathbf{P})$ increase, because $\frac{\partial U(\mathbf{X},\mathbf{P})}{\partial p_{j}^{(2)}}>0$ and $\frac{\partial U(\mathbf{X},\mathbf{P})}{\partial p_{j}^{(4)}}<0$. In addition, the exchange of $\Delta p$ between the two RBs does not break the sum-power constraint. We do this until convergence as in the right figure of Fig. \ref{fig:graph_partial_derivatives}.

Now, the concern is two-fold: 1) consideration of the RBs with zero power and 2) SBSs with $P_{j,\textrm{max}}>\sum_{c\in\mathcal{C}}p_{j}^{(c)}$.

The first concern can be completely resolved if we take $\Delta p$ from the transmission power for the RB with the smallest $\frac{\partial U(\mathbf{X},\mathbf{P})}{\partial p_{j}^{(2)}}$ at each iteration. Suppose that all the transmission power of RB $c$ is taken at the $t$-th iteration. According to our design choice, this means $\frac{\partial U(\mathbf{X},\mathbf{P})}{\partial p_{j}^{(c)}}\le \frac{\partial U(\mathbf{X},\mathbf{P})}{\partial p_{j}^{(c')}}$ for all $c'\in \mathcal{C}\setminus{c}$. At the next iteration, the power exchange shall be continued for the RBs with non-zero transmission power. At convergence, we will have $\frac{\partial U(\mathbf{X},\mathbf{P})}{\partial p_{j}^{(\hat{c})}}\le \frac{\partial U(\mathbf{X},\mathbf{P})}{\partial p_{j}^{(\tilde{c})}}$ and $\left|\frac{\partial U(\mathbf{X},\mathbf{P})}{\partial p_{j}^{(\tilde{c})}}-\xi_j\right|\le\varepsilon$, $\varepsilon>0$, for all $\tilde{c}\in \mathcal{C}_{\textrm{active}}$ and $\hat{c}\in\mathcal{C}_{\textrm{inactive}}$, where $\mathcal{C}_{\textrm{active}} = \left\{ c | p_{j}^{(c)} > 0, c\in \mathcal{C} \right\}$ and $\mathcal{C}_{\textrm{inactive}} = \left\{ c |  p_{j}^{(c)} =0, c\in \mathcal{C} \right\}$. Therefore, the KKT condition for the RB with zero-transmission power, \eqref{eq:KKT2_b}, can be also satisfied.

The second concern can be resolved as follow. After convergence, if all the $\frac{\partial U(\mathbf{X},\mathbf{P})}{\partial p_{j}^{(c)}}$ values for SBS $j$ become identical to a positive constant or zero, our assumption $P_{j,\textrm{max}}=\sum_{c\in\mathcal{C}}p_{j}^{(c)}$ holds true. On the other hand, if $\frac{\partial U(\mathbf{X},\mathbf{P})}{\partial p_{j'}^{(c)}}$ values for SBS $j'$   become identical to a negative constant, the full transmission power for SBS $j'$ cannot be assumed at the local optimal point. Instead, we should have  $P_{j',\textrm{max}}>\sum_{c\in\mathcal{C}}p_{j'}^{(c)}$. Then, we can repeat the algorithm modifying the total transmission power as $\sum_{c\in\mathcal{C}}p_{j'}^{(c)}=P_{j',\textrm{max}}-\Delta p_{\textrm{total}}$ for a step size $\Delta p_{\textrm{total}}>0$.

In what follows, detailed parameters optimization are presented, followed by the overall proposed algorithm.

\subsubsection{Selection of the Two RBs for the Transmission Power Exchange\label{sec:exchange}}
Let us update $p_{j}^{(c_1)}$ and $p_{j}^{(c_2)}$ at the $t$-th iteration in design of the transmission power for SBS $j$. The updated power is denoted as $\mathbf{P}_{[t]}\in\mathbb{R}^{B\times C}$, where $ \left[ \mathbf{P}_{[t]}\right]_{jc}=p_{j,[t]}^{(c)}$ is defined by
\begin{equation} \label{eq:p_jt}
    p_{j,[t]}^{(c)} = \begin{cases}
    p_{j,[t-1]}^{(c)}+\Delta p_{j,[t]}, & \textrm{if}~c=c_1,\\
    p_{j,[t-1]}^{(c)}-\Delta p_{j,[t]}, & \textrm{if}~c=c_2,\\
    p_{j,[t-1]}^{(c)}, & \textrm{otherwise}.
    \end{cases}
\end{equation}
Here, $\Delta p_{j,[t]}$ denotes the amount of transmission power exchange between the selected RBs at the $t$-th design iteration in SBS $j$.
Then, the proportional fairness with $\mathbf{P}_{[t]}$ is approximated for small $\Delta p_{j,[t]}$ by Taylor series with respect to $\Delta p_{j,[t]}$ as follow:
\begin{equation}\label{eq:Approx_U}
   U(\mathbf{X},\mathbf{P}_{[t]})= U(\mathbf{X},\mathbf{P}_{[t-1]}) + \left.\left(\frac{\partial U(\mathbf{X},\mathbf{P})}{\partial p_{j}^{(c_1)}}-\frac{\partial U(\mathbf{X},\mathbf{P})}{\partial p_{j}^{(c_2)}}\right)\right|_{\mathbf{P} = \mathbf{P}_{[t-1]}} \Delta p_{j,[t]} + O({\Delta p_{j,[t]}} ^2).
\end{equation}
In order to maximize \eqref{eq:Approx_U} for given $\Delta p_{j,[t]}$, the two RBs $c_1$ and $c_2$ are chosen by
\begin{equation}
c_{1} =\argmax_{c\in\mathcal{C}} \left.\frac{\partial U}{\partial p_{j}^{(c)}}\right|_{\mathbf{P}=\mathbf{P}_{[t-1]}} \text{and }\hspace{10pt} c_{2} =\argmin_{c\in\mathcal{C},\hspace{1pt}p_{j,[t-1]}^{(c)}\ne 0} \left.\frac{\partial U}{\partial p_{j}^{(c)}}\right|_{\mathbf{P}=\mathbf{P}_{[t-1]}}.
\label{eq:RB Selecting}
\end{equation}

\subsubsection{Design of the Power Exchange\label{sec:Power_design}}

For given RB indices $(c_1)$ and $(c_2)$ for SBS $j$,
$\Delta p_{j,[t]}$ is designed to satisfy the optimality condition $\frac{\partial U(\mathbf{X},\mathbf{P})}{\partial p_{j}^{(c_1)}}=\frac{\partial U(\mathbf{X},\mathbf{P})}{\partial p_{j}^{(c_2)}}$. Let us denote the difference in the partial derivatives at the $t$-th iteration as
\begin{equation}
    f(\mathbf{P}_{[t]}) = f_1(\mathbf{P}_{[t]})-f_2(\mathbf{P}_{[t]}),
\end{equation}
where $f_{i}(\mathbf{P}_{[t]})= \left.\frac{\partial U(\mathbf{X},\mathbf{P})}{\partial p_{j}^{(c_i)}}\right|_{\mathbf{P}=\mathbf{P}_{[t]}}$ is defined over $i\in\{1,2\}$.
The following proposition establishes the approximate of the optimal $\Delta p_{j,[t]}$ in a closed-form.
\begin{proposition}\label{prop:delta_p}
For any given RB $c_1$ and $c_2$ of SBS $j$, $\Delta p_{j,[t]}$ leading to $f(\mathbf{P}_{[t]})= O\left( {\Delta p_{j,[t]}}^3 \right)$ is obtained by
\begin{equation}\label{eq:opt_delta}
\Delta p_{j,[t]}^*=\frac{-f'(\mathbf{P}_{[t-1]})+\mathrm{sgn}\left[f'(\mathbf{P}_{[t-1]})\right]\sqrt{f'(\mathbf{P}_{[t-1]})^{2}-2f''(\mathbf{P}_{[t-1]})f(\mathbf{P}_{[t-1]})}}{f''(\mathbf{P}_{[t-1]})},
\end{equation}
where $\mathrm{sgn}[\cdot]$ denotes the sign function,  $f''(\mathbf{P}_{[t-1]})=f_1^{(2)}(\mathbf{P}_{[t-1]})-f_2^{(2)}(\mathbf{P}_{[t-1]})$, and $f'(\mathbf{P}_{[t-1]}) =f_1^{(1)}(\mathbf{P}_{[t-1]})-f_2^{(1)}(\mathbf{P}_{[t-1]})
$. Here, with the assumption $\lambda_i\gg W$, the funtions are given by
\begin{align}
\label{eq:partial_2}
f_{i}(\mathbf{P}_{[t-1]})=\left.\frac{\partial U(\mathbf{X},\mathbf{P})}{\partial p_{j}^{(c_i)}}\right|_{\mathbf{P}=\mathbf{P}_{[t-1]}} & =\sum_{k\in\mathcal{B}}\frac{S}{\lambda_{i^{(k,c_i)}}\ln2}-\sum_{k\neq j}\frac{Q}{\lambda_{i^{(k,c_i)}}\ln2},
\\
\label{eq:partial_3}
f_{i}^{(1)}(\mathbf{P}_{[t-1]})=\left.\frac{\partial f_i(\mathbf{P}_{[t]})}{\partial \Delta p_{j,[t]}}\right|_{\Delta p_{j,[t]}=0} & =\sum_{k\in\mathcal{B}}\frac{(-1)^iS^{2}}{\lambda_{i^{(k,c_i)}}\ln2} -\sum_{k\neq j}\frac{(-1)^iQ^{2}}{\lambda_{i^{(k,c_i)}}\ln2},
\\
\label{eq:partial_4}
f_{i}^{(2)}(\mathbf{P}_{[t-1]})=\left.\frac{\partial^{2}f_i(\mathbf{P}_{[t]})}{\partial \Delta p_{j,[t]}^{2}}\right|_{\Delta p_{j,[t]}=0} & =\sum_{k\in\mathcal{B}}\frac{2S^{3}}{\lambda_{i^{(k,c_i)}}\ln2}
 -\sum_{k\neq j}\frac{2Q^{3}}{\lambda_{i^{(k,c_i)}}\ln2},
\end{align}
where $S=W\left[\frac{\mathrm{SNR}_{i^{(k,c_i)}k}^{(c_i)}}{1+\sum_{b\in\mathcal{B}}\mathrm{SNR}_{i^{(b,c_i)}b}^{(c_i)}p_{b,[t-1]}^{(c_i)}}\right]$, $Q=W\left[\frac{\mathrm{SNR}_{i^{(k,c_i)}k}^{(c_i)}}{1+\sum_{b\neq k}\mathrm{SNR}_{i^{(b,c_i)}b}^{(c_i)}p_{b,[t-1]}^{(c_i)}}\right]$,
$i^{(k,c)}$ denotes the user served by SBS $k$ on RB $c$, $\mathrm{SNR}_{ij}^{(c)}=\frac{|h_{ij}^{(c)}|^{2}}{\sigma^{2}}$,
$i^{(k,c)}=\argmax_{i'\in\mathcal{N}}x_{i'k}^{(c)}$, and  $\lambda_{i^{(k,c)}}=\sum_{j\in\mathcal{B}}\sum_{l\in\mathcal{C}}R_{i^{(k,c)}j}^{(l)}x_{i^{(k,c)}j}^{(l)}$.
\end{proposition}
\begin{IEEEproof}
Proof is in Appendix \ref{sec:Appendix_C}.
\end{IEEEproof}
According to Proposition \ref{prop:delta_p}, $\Delta p_{j,[t]}^*$ chosen as \eqref{eq:opt_delta} gives us the KKT optimality $f(\mathbf{P}_{[t]})\rightarrow 0$ as iteration grows, where $\Delta p_{j,[t]}<1$. On the other hand, however, the backhaul constraint \eqref{eq:P3_c} should be taken into consideration by limiting the maximum possible value of $\Delta p_{j,[t]}$, as shown in Proposition \ref{prop:Given-remained-data}.
\begin{proposition}
\label{prop:Given-remained-data} For given $\mathbf{X}$, the maximum possible $\Delta p_{j,[t]}$ that satisfies the backhaul capacity is obtained in terms of  the remaining
backhaul capacity of SBSs as
\begin{equation}\label{eq:limit_delta_p}
\Delta p_{j,[t]}\leq\min\left\{ \min_{k\in\mathcal{B}\setminus\{j\}}\left\{ p_{k,[t-1]}^{(c_{2})}\left(1-2^{\frac{-L_{k}(\mathbf{X},\mathbf{P}_{[t-1]})}{W}}\right)\right\} ,p_{j,[t-1]}^{(c_{1})}\left(2^{\frac{L_{j}(\mathbf{X},\mathbf{P}_{[t-1]})}{W}}-1\right)\right\},
\end{equation}
where $L_j(\mathbf{X},\mathbf{P}_{[t]})=Z_j-\left.\sum_{i\in\mathcal{N}}\sum_{c\in\mathcal{C}}R_{ij}^{(c)}x_{ij}^{(c)}\right|_{\mathbf{P}=\mathbf{P}_{[t]}}$.
\end{proposition}
\begin{IEEEproof}
shown in Appendix \ref{sec:Appendix_D}.
\end{IEEEproof}
From the Proposition \ref{prop:Given-remained-data}, the power exchange with backhaul consideration $\Delta{p}_{j,[t]}$ is denoted as
\begin{equation}\label{eq:delta_p_jt}
\Delta{p}_{j,[t]}=\min\left\{ \min_{k\in\mathcal{B}\setminus\{j\}}\left\{ p_{j,[t-1]}^{(c_{2})}\left(1-2^{\frac{-L_{k}(\mathbf{X},\mathbf{P}_{[t-1]})}{W}}\right)\right\} ,p_{j,[t-1]}^{(c_{1})}\left(2^{\frac{L_{j}(\mathbf{X},\mathbf{P}_{[t-1]})}{W}}-1\right),\Delta p_{j,[t]}^*\right\}.
\end{equation}


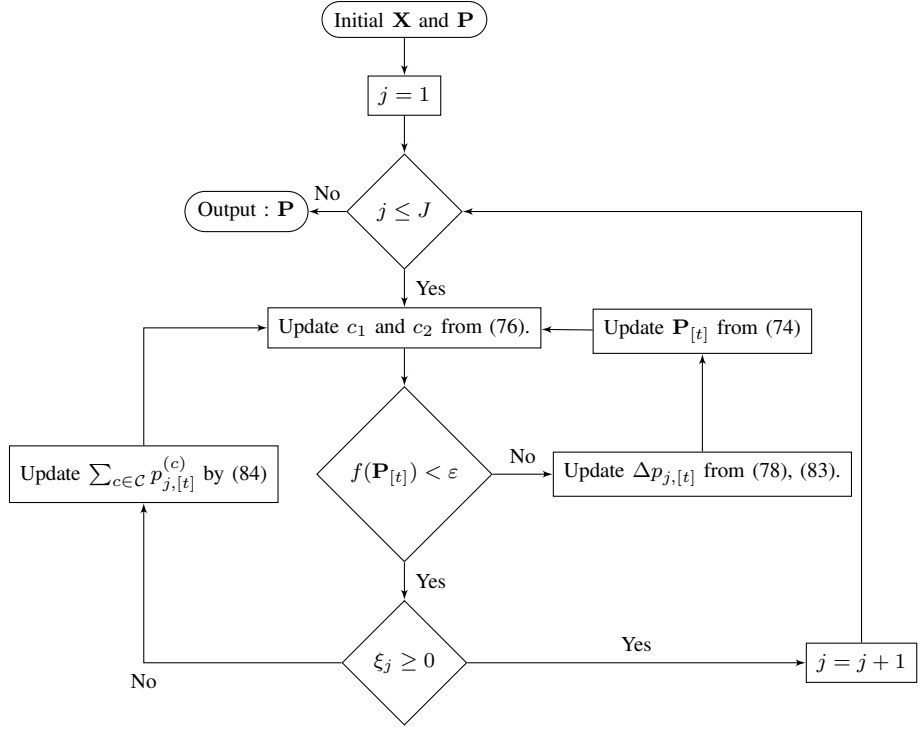
\begin{figure}
\begin{tikzpicture}{node distance = 1.5cm, auto}
    \tikzstyle{every node}=[font=\scriptsize]
    \tikzstyle{terminal} = [rounded rectangle, text centered, draw=black]
    \tikzstyle{process} = [rectangle, text centered, draw=black]
    \tikzstyle{decision} = [diamond, text centered, draw=black]
    \tikzstyle{line} = [draw, -latex']
\node [terminal] (init) {Initial $\mathbf{X}$ and $\mathbf{P}$};
\node [process, below = 0.5 of init] (jeq1) {$j=1$};
\node [decision, below = 0.5 of jeq1] (jleJ) {$j\le J$};
\node [terminal, left = 0.5 of jleJ] (output) {Output : $\mathbf{P}$};
\node [process, below = 0.5 of jleJ] (c_1) {Update $c_1$ and $c_2$ from \eqref{eq:RB Selecting}.
};
\node [decision, below = 0.5 of c_1] (dec) {$f(\mathbf{P}_{[t]})<\varepsilon$};
\node [process, right = 0.8 of dec] (delta_p) {Update $\Delta p_{j,[t]}$ from \eqref{eq:opt_delta}, \eqref{eq:delta_p_jt}.};
\node [process, above = 1.3 of delta_p] (Pt) {Update $\mathbf{P}_{[t]}$ from \eqref{eq:p_jt}};
\node [decision, below = 0.5 of dec] (xi) {$\xi_j\ge 0$};
\node [process, left = 0.5 of dec] (total_p) {Update $\sum_{c\in\mathcal{C}}p_{j,[t]}^{(c)}$ by \eqref{eq:sum_power_control}};
\node [process, right = 4.5 of xi] (jeqj1) {$j=j+1$};
\path[line] (init)--(jeq1);
\path[line] (jeq1)--(jleJ);
\path[line] (jleJ)--node[anchor=west]{Yes}(c_1);
\path[line] (c_1)--(dec);
\path[line] (dec)--node[anchor=west]{Yes}(xi);
\path[line] (xi)--node[anchor=south]{Yes}(jeqj1);
\path[line] (jeqj1)|-(jleJ);
\path[line] (dec)--node[anchor=south]{No}(delta_p);
\path[line] (xi)-|node[anchor=north]{No}(total_p);
\path[line] (Pt)--(c_1);
\path[line] (delta_p)--(Pt);
\path[line] (total_p)|-(c_1);
\path[line] (jleJ)-- node[anchor=south]{No} (output);
\end{tikzpicture}\centering\caption{Flow chart of the proposed power control algorithm\label{fig:Flow_chart}}
\end{figure}

Then, we may have $\mathbf{P}_{[t]}$ that satisfies the KKT condition \eqref{eq:KKT2} except \eqref{eq:KKT2_e} with the proposed algorithm; that is,  $\frac{\partial U(\mathbf{X},\mathbf{P})}{\partial p_j^{(\hat{c})}}\leq \xi_j$ and $\frac{\partial U(\mathbf{X},\mathbf{P})}{\partial p_j^{(\tilde{c})}}= \xi_j$ for all $j\in\mathcal{B},\tilde{c}\in\mathcal{C}_{\textrm{active}},\hat{c}\in\mathcal{C}_{\textrm{inactive}}$, if the backhaul constraint is irrelevant, i.e., $Z_j-\sum_{i\in\mathcal{N}}\sum_{c\in\mathcal{C}}R_{ij}^{(c)}x_{ij}^{(c)}$ is relatively large.

\subsubsection{Design of $\Delta p_{total}$}
If the backhaul constraint is very tight, the KKT condition \eqref{eq:KKT2} cannot be satisfied with the current transmission power assumption. In such case, although the proposed algorithm ends up with the same conditions, $\xi_j$ is negative, which contradicts to \eqref{eq:KKT2_e}.  At convergence after repeating Section \ref{sec:exchange} and \ref{sec:Power_design} at $t=T_n$, if $\xi_j<0$, then the sum-power should be reduced to satisfy the KKT conditions \eqref{eq:KKT2}.
Since the gradient of the Lagrangian $L(\mathbf{P},\bm{\xi},\bm{\varphi})$ in \eqref{eq:L_p} is $\xi_j$, the gradient method yields the updated sum-power constraint for SBS $j$ with $\xi_j<0$ being $\sum_{c\in\mathcal{C}}p_{j,[T_n+1]}^{(c)}=\sum_{c\in\mathcal{C}}p_{j,[T_n]}^{(c)}+\Delta p_{\textrm{total}}^{(n)}$, which denoted by
\begin{equation}\label{eq:sum_power_control}
\Delta p_{\textrm{total}}^{(n)} = \min\{\gamma \xi_j,P_{j,\textrm{max}}-\sum_{c\in\mathcal{C}}p_{j,[T_n]}^{(c)}\},
\end{equation}
where $\gamma$ denotes step size.

\subsection{Overall Algorithm}
The overall proposed PC algorithm is summarized in Fig. \ref{fig:Flow_chart}.
\subsection{Complexity Analysis}

\begin{table}
\caption{Complexity comparison in flops}
\begin{tabular}{|c|c|c|}
\hline
Algorithm & UA and RA & PC\tabularnewline
\hline
energy-constrained FFRA \cite{Q_Han15} &$O\left(\frac{JN}{\epsilon^{2}}\right)$ & - \tabularnewline
\hline
modified unconstrained FFRA \cite{Kaiming_JSAC}& $O((I_{\textrm{SBS}})JN)$ &$O((I_{\textrm{SBS}})J^{2})$
\tabularnewline
\hline
proposed algorithm&$O\left(I_{\zeta}I_{\nu}({NC})\right)$&$O\left(I_{\mathbf{P}}(J^{2}+J^2C)\right)$\tabularnewline
\hline
genetic algorithm & \multicolumn{2}{c|}{$O(G_{\textrm{max}}M_{p}J^2NC^2)$}
\tabularnewline
\hline
global optimal solution&$O\left(N^{BC}\right)$&-
\tabularnewline
\hline
\end{tabular}\centering\label{table:table_1}
\end{table}

In this subsection, the computational complexity in flops for the proposed RA algorithm
based on Algorithm 2 and PC algorithm based on Fig. \ref{fig:Flow_chart} is analyzed in comparison to the existing approaches in Table \ref{table:table_1}. For comparison, the energy-constrained fractional frequency RA (FFRA) algorithm \cite{Q_Han15} and unconstrained FFRA algorithm \cite{Kaiming_JSAC} are considered, where the RA problem is relaxed to a continuous optimization problem with real-valued variables. For fair comparison, we have modified the unconstriant FFRA algorithm to consider the backhaul constraint and PC.

The
overall proposed algorithm has computational complexity of $O\left(I_{\bm{\zeta}}I_{\bm{\nu}}(NC)+I_{\mathbf{P}}(J^2+J^2C)\right)$, where $I_{\bm{\zeta}}$, $I_{\mathbf{P}}$, and $I_{\bm{\nu}}$ denote the numbers of iterations needed for $\bm{\zeta}$, $\mathbf{P}$, and $\bm{\nu}$ to converge,
 respectively. The global optimal solution of the problem \eqref{eq:P_UA} requires exponential computational complexity of $O\left(N^{BC}\right)$.  The energy-constrained FFRA has the computational complexity of $O(\frac{JN}{\epsilon^{2}})$ overall, where $\epsilon$ denotes the convergence threshold. The computational complexity of the modified unconstrained FFRA is $O((I_{\textrm{SBS}})\mathcal{BN})+O((I_{\textrm{SBS}})\mathcal{B}^{2})$, where $I_{\textrm{SBS}}$ denotes the number of iterations. The overall computational complexity of the genetic algorithm is $O(G_{\textrm{max}}M_{p}J^2NC^2)$, where $G_{\textrm{max}}$ and $M_p$ denote the maximum generation and the population size, respectively. Numerical results for the computational complexity shall be presented in Section \ref{sec:Numerical-Results}.

\pagebreak[0]
\section{Numerical Results\label{sec:Numerical-Results}}
\begin{table}
\caption{Simulation Parameters\label{tab:Parameters}}
\begin{tabular}{|c|c|}
\hline
Network model & 3GPP scenario 2a \cite{3gpp_36_872}\tabularnewline
\hline
Number of small cells & 4\tabularnewline
\hline
Number of users in the cluster & 40\tabularnewline
\hline
Number of resource blocks & 100\tabularnewline
\hline
$W$(kHz) & 180\tabularnewline
\hline
Backhaul capacity(Mbps) & 20-100\tabularnewline
\hline
Transmission power(dBm) & 35\tabularnewline
\hline
Noise power(dBm/Hz) & -174\tabularnewline
\hline
Cluster diameter(m) & 1000\tabularnewline
\hline
Pathloss(dB) & $38+30\log_{10}(d)$\tabularnewline
\hline
\end{tabular}\centering
\end{table}

\pagebreak[0]The proportional fairness of the proposed scheme is numerically evaluated in the 3GPP small cell scenario 2a. That is, a small cell is interfered only by other small cells, not macro cells \cite{3gpp_36_872}. The system parameters in table \ref{tab:Parameters}
are based on \cite{3gpp_36_872,3gpp_36_874}, which are used for the simulations.

\begin{figure}[t]
\begin{minipage}[b]{.48\textwidth}
  \centering
  \includegraphics[width=\textwidth]{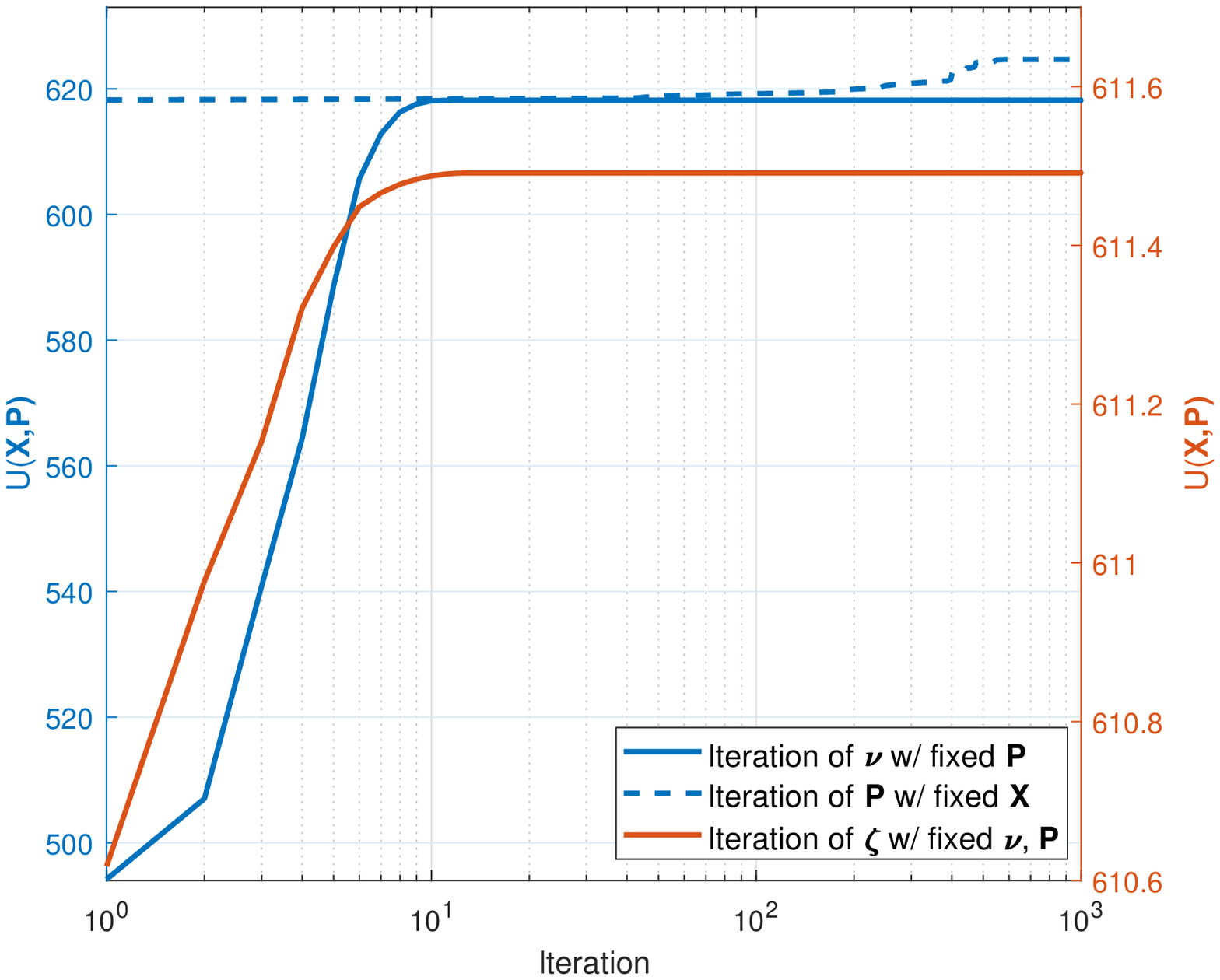}
  \caption{Proportional fairness versus the number of iterations for $\bm{\zeta}$, $\bm{\nu}$ and $\mathbf{P}$ with backhaul capacity of 81Mbps}
  \label{fig:g_graph}
\end{minipage}
\hfill
\begin{minipage}[b]{.48\textwidth}
  \centering
  \includegraphics[width=\textwidth]{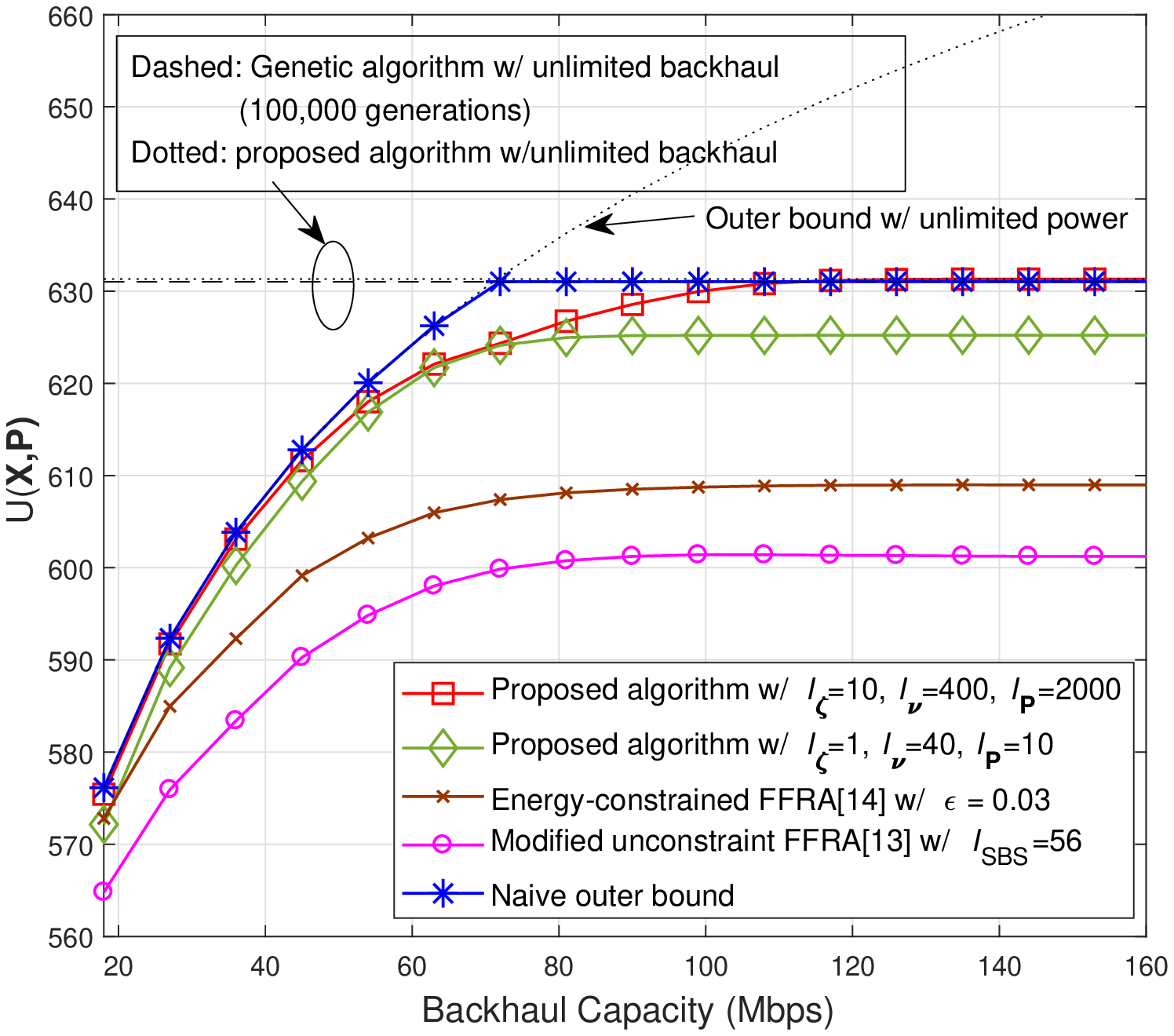}\centering
  \caption{Proportional fairness versus backhaul capacity \label{fig:Network-utility-versus}}
\end{minipage}
\end{figure}

The proportional fairness $U(\mathbf{X},\mathbf{P})$ of the proposed scheme, i.e. UA and RA in Algorithm \ref{alg:Cyclic coordinate descent} and PC in Fig. \ref{fig:Flow_chart}, versus the number of iterations is depicted in Fig. \ref{fig:g_graph} in order to show the convergence of $\bm{\zeta}$, $\bm{\nu}$ and $\mathbf{P}$. As shown in the figure, $\bm{\zeta}$ converges in Algorithm \ref{alg:Cyclic coordinate descent} within a reasonable number of iterations, which is usually less than 10. Though, the iterations needed for convergence
of $\bm{\nu}$ and $\mathbf{P}$ are around 400 and 2000, respectively, only a few iterations also give us relatively high performance.

In Fig. \ref{fig:Network-utility-versus}, the proportional fairness is shown with respect to the backhaul capacity. The proposed scheme is implemented with the high complexity ($I_{\bm{\zeta}}=10$, $I_{\bm{\nu}}=400$, $I_{P}=2000$) and low complexity ($I_{\bm{\zeta}}=1$, $I_{\bm{\nu}}=40$, $I_{P}=10$) settings, requiring orders of 176160 and $2\cdot10^{7}$ flops, respectively. On the other hand, the energy-constrained FFRA with $\epsilon=0.03$ \cite{Q_Han15} and modified unconstrained FFRA with $I_{\textrm{SBS}}=56$ \cite{Kaiming_JSAC} require the overall complexity of orders of $177777$ and $9856$ flops, respectively. It is shown that the proposed scheme with both of the settings significantly outperform the previous approaches, obtaining higher frequency diversity gain due to finer per-RB UA, RA, and PC. Particularly, the proposed scheme with the low complexity setting achieves higher proportional fairness than the energy-constrained FFRA even with lower computational complexity for the backhaul capacity higher than 20Mbps.

For comparison, the outer bound without power constraint is considered, which can be derived using the inequality/equality for arithmetic and geometric averages, as shown in Proposition 1 of \cite{Q_Ye_TWC}. The proposed scheme asymptotically achieves this outer bound as the backhaul capacity becomes small, where the power constraints become satisfied in the derivation of the outer bound, yielding a tight upper bound. On the other hand, in persuit of finding the global optimal solution, we also consider the genetic algorithm but without backhaul capacity. For the genetic algorithm, we considered $100$ population
, $0.8$ crossover fraction, $10^{5}$ maximum generations, and $50$ elite counts. For convergence of one simulation environment, the genetic algorithm required about 5 hours whereas the proposed scheme with the high complexity convereged within 1 min on average. As the backhaul capacity increases, the proposed scheme asymptotically achieves the performance of the genetic algorithm without power constraint with much lower computational complexity.


\section{Conclusions\label{sec:Conclusions}}
We have considered a UA, RA, and PC problem with limited backhaul capacity to maximize the proportional fairness of cooperative multicel networks .
We have proposed a cascaded iterative algorithm to solve the problem and show the achievability of the optimal solution. The simulation results have shown that the proposed scheme closely achieves the globally optimal proportional fairness, which can be obtained with exponential computational complexity, at the cost of reasonably increased computational complexity compared with the existing schemes.
\begin{appendices}

\section{Proof of Lemma \ref{lem:gap_bound}}\label{sec:Appendix_gap}
Let $\mathbf{X}^*$ be a solution from Algorithm \ref{alg:Cyclic coordinate descent} and $\epsilon_j=Z_j-\sum_{i\in\mathcal{N}}\sum_{c\in\mathcal{C}}R_{ij}^{(c)}x_{ij}^{(c)*}$. Then, for any $\mathbf{X}^\prime\in A(\mathbf{X}^*)$ in \eqref{eq:2_dist_point} that satisfies the constraint of backhaul capacity \eqref{eq:backhaul_ineq}, we have $\epsilon_j\geq0$ for all $j\in\mathcal{B}$ and
 \begin{equation}\label{eq:appendix_A_ineq}
 \sum_{i\in\mathcal{N}}R_{ij}^{(c)}\left(\zeta_{ij}^{(c)}(\mathbf{X}^*)-\nu_{j}\right){x_{ij}^{(c)}}^* \geq \sum_{i\in\mathcal{N}}R_{ij}^{(c)}\left(\zeta_{ij}^{(c)}(\mathbf{X}^*)-\nu_{j}\right){x_{ij}^{(c)}}^\prime, \forall (j,c)\in\mathcal{B}\times\mathcal{C}, \hspace{5pt}\forall \mathbf{X'}\in A(\mathbf{X}^*),
 \end{equation}
 From \eqref{eq:appendix_A_ineq}, we get
 \begin{equation}\label{eq:ineq_appendix1}
 \sum_{i\in\mathcal{N}}R_{ij}^{(c)}\zeta_{ij}^{(c)}(\mathbf{X}^*){x_{ij}^{(c)}}^\prime- R_{ij}^{(c)}\zeta_{ij}^{(c)}(\mathbf{X}^*){x_{ij}^{(c)}}^*\leq \sum_{i\in\mathcal{N}}\nu_{j}R_{ij}^{(c)}({x_{ij}^{(c)}}^\prime-{x_{ij}^{(c)}}^*) \leq \epsilon_j\nu_j,\forall (j,c)\in\mathcal{B}\times\mathcal{C},\forall \mathbf{X}'\in A(\mathbf{X}^*),
 \end{equation}
Let $\mathbf{X}$ and $\mathbf{X}'$ be different for only $(\hat{j},\hat{c})$. Then, first term of \eqref{eq:ineq_appendix1} is bounded by  \linebreak[0] \pagebreak[0]
\begin{align}
 \sum_{i\in\mathcal{N}}R_{i\hat{j}}^{(\hat{c})}\zeta_{i\hat{j}}^{(\hat{c})}(\mathbf{X}^*){x_{i\hat{j}}^{(\hat{c})}}^\prime- R_{i\hat{j}}^{(\hat{c})}\zeta_{i\hat{j}}^{(\hat{c})}(\mathbf{X}^*){x_{i\hat{j}}^{(\hat{c})}}^*
 &=\frac{R_{T(\mathbf{X}',\hat{j},\hat{c})\hat{j}}^{(\hat{c})}}{\left.V_{T(\mathbf{X}',\hat{j},\hat{c})}(\mathbf{X}')\right|_{{x_{i\hat{j}}^{(\hat{c})}}^\prime=0, ~\forall i\in\mathcal{N}}}-\frac{R_{T(\mathbf{X}^*,\hat{j},\hat{c})\hat{j}}^{(\hat{c})}}{\left.V_{T(\mathbf{X}^*,\hat{j},\hat{c})}(\mathbf{X}^*)\right|_{{x_{i\hat{j}}^{(\hat{c})}}^*=0, ~\forall i\in\mathcal{N}}} \linebreak[0]\pagebreak[0]\nonumber\\ &=\frac{V_{T(\mathbf{X}',\hat{j},\hat{c})}(\mathbf{X}')}{\left.V_{T(\mathbf{X}',\hat{j},\hat{c})}(\mathbf{X}')\right|_{{x_{i\hat{j}}^{(\hat{c})}}^\prime=0, ~\forall i\in\mathcal{N}}}-\frac{V_{T(\mathbf{X}^*,\hat{j},\hat{c})}(\mathbf{X}^*)}{\left.V_{T(\mathbf{X}^*,\hat{j},\hat{c})}(\mathbf{X}^*)\right|_{{x_{i\hat{j}}^{(\hat{c})}}^*=0, ~\forall i\in\mathcal{N}}} \linebreak[0]\pagebreak[0]\nonumber\\
 &\geq\frac{\prod_{i\in\mathcal{N}}V_{i}(\mathbf{X}^\prime)-\prod_{i\in\mathcal{N}}V_{i}(\mathbf{X}^*)}{\prod_{i\in\mathcal{N}}V_i(\mathbf{X}^*)} =\frac{e^{U(\mathbf{X}',\mathbf{P})}}{e^{U(\mathbf{X}^*,\mathbf{P})}}-1\linebreak[0] \pagebreak[0]\label{eq:appendA_ineq}
 \end{align}
 \pagebreak[0]where $T(\mathbf{X},j,c) = \argmax_{i\in\mathcal{N}}x_{ij}^{(c)}$ and $V_i(\mathbf{X})=\sum_{j\in\mathcal{B}}\sum_{c\in\mathcal{C}}R_{ij}^{(c)}x_{ij}^{(c)}$.  From \eqref{eq:ineq_appendix1} and \eqref{eq:appendA_ineq}, because $\hat{j}$ is determined by $\mathbf{X}'$, we have
\begin{equation}
U(\mathbf{X}',\mathbf{P})-U(\mathbf{X}^*,\mathbf{P})\leq \log(1+\epsilon_{\hat{j}}\nu_{\hat{j}})\leq\max_{j\in\mathcal{B}}\log(1+\epsilon_j\nu_j)\leq\max_{j\in\mathcal{B}}\epsilon_j\nu_j, \hspace{5pt}\forall \mathbf{X}'\in A(\mathbf{X}^*).
\end{equation} Then, the proportional fairness gap between the solution $\mathbf{X}^*$ and $\mathbf{X}'\in A(\mathbf{X}^*)$ is bounded by $\max_{j\in\mathcal{B}}\epsilon_j\nu_j$. \\
In the Theorem \ref{thm:global}, $Z_j\neq \sum_{i\in\mathcal{N}}\sum_{c\in\mathcal{C}}R_{ij}^{(c)}x_{ij}^{(c)}$ with probability $1$ since $R_{ij}^{(c)}$ is a random variable and the dimension of $R_{ij}^{(c)}$ is finite. As $C\rightarrow\infty$, the dimension of  $R_{ij}^{(c)}$ is infinite. Then, there exists $\mathbf{X}$ such that $Z_j= \sum_{i\in\mathcal{N}}\sum_{c\in\mathcal{C}}R_{ij}^{(c)}x_{ij}^{(c)}$ with probability $1$.

\section{Proof of Lemma \ref{lem:KKT_power} \label{sec:Appendix_B}}
The KKT condition obtained from the lagrangian \eqref{eq:L_p} is denoted as
\begin{subnumcases}{}\label{eq:KKT}
\frac{\partial U(\mathbf{X},\mathbf{P})}{\partial p_{j}^{(c)}}-\xi_{j}+\varphi_{j}^{(c)}=0, \hspace{5pt}\forall j\in\mathcal{B}, c\in\mathcal{C}\label{eq:KKT_a}\\
\xi_{j}\left(P_{j,\textrm{max}}-\sum_{c\in\mathcal{C}}p_{j}^{(c)}\right)=0, \hspace{5pt}\forall j\in\mathcal{B}\label{eq:KKT_b}\\
\varphi_{j}^{(c)}p_{j}^{(c)}=0, \hspace{5pt} \forall j\in\mathcal{B}, c\in\mathcal{C}\label{eq:KKT_c}\\
P_{j,\textrm{max}}-\sum_{c\in\mathcal{C}}p_{j}^{(c)}\geq 0, \hspace{5pt} \forall j\in \mathcal{B}\label{eq:KKT_d}\\
\varphi_{j}^{(c)} \geq 0, \hspace{5pt} \forall j\in\mathcal{B}, c\in\mathcal{C}.\label{eq:KKT_e}
\end{subnumcases}
In addition, $\mathbf{P}$ should satisfy the constraints \eqref{eq:P4_b} and \eqref{eq:P4_c} of the problem \eqref{eq:problem_power_simple}. From \eqref{eq:KKT_a}, \eqref{eq:KKT_b} and \eqref{eq:KKT_c}, we have $\frac{\partial U(\mathbf{X},\mathbf{P})}{\partial p_{j}^{(c)}} = \xi_j -\varphi_j^{(c)}$, where $\xi_{j} =0 $ $\text{if }\sum_{c\in\mathcal{C}}p_{j}^{(c)}<P_{j,\textrm{max}}$ and $\varphi_{j}^{(c)} = 0$ $\text{if } p_{j}^{(c)}>0$. Then, the partial derivatives $\frac{\partial U(\mathbf{X},\mathbf{P})}{\partial p_{j}^{(c)}}$ is represented in terms of $\xi_j$ and $\varphi_j^{(c)}$ as
\begin{equation}\label{eq:partial_U_2}
\frac{\partial U(\mathbf{X},\mathbf{P})}{\partial p_{j}^{(c)}}=
\begin{cases}
\xi_{j}, \hspace{5pt} &\text{if} \hspace{5pt} p_{j}^{(c)} >0,\hspace{5pt} \sum_{c\in\mathcal{C}}p_{j}^{(c)}= P_{j,\textrm{max}},\\
\xi_{j}-\varphi_{j}^{(c)}, \hspace{5pt} &\text{if} \hspace{5pt} p_{j}^{(c)}=0,\hspace{5pt} \sum_{c\in\mathcal{C}}p_{j}^{(c)}= P_{j,\textrm{max}},\\
0, \hspace{5pt} &\text{if} \hspace{5pt} p_{j}^{(c)} >0,\hspace{5pt} \sum_{c\in\mathcal{C}}p_{j}^{(c)}< P_{j,\textrm{max}},\\
-\varphi_{j}^{(c)}, \hspace{5pt} &\text{if} \hspace{5pt} p_{j}^{(c)} =0,\hspace{5pt} \sum_{c\in\mathcal{C}}p_{j}^{(c)}< P_{j,\textrm{max}}.\\
\end{cases}
\end{equation}
From \eqref{eq:partial_U_2}, the KKT condition \eqref{eq:KKT} is represented by \eqref{eq:KKT2}.

\pagebreak[0]
\section{Proof of Proposition \ref{prop:delta_p}\label{sec:Appendix_C}}
\pagebreak[0]The second-order approximated $f_{i}(\mathbf{P}_{[t]})$ from Taylor series with small $\Delta p_{j,[t]}$ is written as
\begin{equation}\label{eq:f_1,f_2}
f_{i}(\mathbf{P}_{[t]})  =\left.\frac{\partial U(\mathbf{X},\mathbf{P})}{\partial p_{j}^{(c_{i})}}\right|_{\mathbf{P}=\mathbf{P}_{[t]}}
= f_i^{(2)}(\mathbf{P}_{[t-1]})\frac{ {\Delta p_{j,[t]}}^{2}}{2!}+f_{i}^{(1)}(\mathbf{P}_{[t-1]})\Delta p_{j,[t]}+f_i(\mathbf{P}_{[t-1]})+O\left(\Delta p_{j,[t]}^3\right),
\end{equation}
where $f_{i}^{(n)}(\mathbf{P}_{[t-1]}) =  \left.\frac{\partial^n f_i(\mathbf{P}_{[t]})}{\partial {\Delta p_{j,[t]}}^n}\right|_{\Delta p_{j,[t]}=0}$.
From \eqref{eq:f_1,f_2}, we have
\begin{equation}
f(\mathbf{P}_{[t]})=f_1(\mathbf{P}_{[t]})-f_2(\mathbf{P}_{[t]})= \frac{f''(\mathbf{P}_{[t-1]})}{2!}{(\Delta p_{[t]})}^{2}+f'(\mathbf{P}_{[t-1]})\Delta p_{j,[t]
}+f(\mathbf{P}_{[t-1]})+O\left(\Delta p_{j,[t]}^3\right),\label{eq:second_order}
\end{equation}
where $f''(\mathbf{P}_{[t-1]})=f_1^{(2)}(\mathbf{P}_{[t-1]})-f_2^{(2)}(\mathbf{P}_{[t-1]})$ and $f'(\mathbf{P}_{[t-1]}) =f_1^{(1)}(\mathbf{P}_{t-1})-f_2^{(1)}(\mathbf{P}_{t-1})$.
Then, the solution of $f(\mathbf{P}_{[t]})=O\left(\Delta p_{j,[t]}^3\right)$ is denoted as
\begin{equation}\label{eq:delta_p_pm}
\Delta p_{j,[t]}=\frac{-f'(\mathbf{P}_{[t-1]})\pm\sqrt{\left\{f'(\mathbf{P}_{[t-1]})\right\}^2-2f''(\mathbf{P}_{[t-1]})f(\mathbf{P}_{[t-1]})}}{f''(\mathbf{P}_{[t-1]})}.
\end{equation}
Because the aim is to obtain $\Delta p_{j,[t]}$ such that $f(\mathbf{P}_{[t]})=O\left(\Delta p_{j,[t]}^3\right)$, $\Delta p_{j,[t]}=0$ in \eqref{eq:delta_p_pm} if $f(\mathbf{P}_{[t-1]})=O\left(\Delta p_{j,[t]}^3\right)$.
To this end, the sign of the square-root term should be  $\mathrm{sgn}\left[f'(\mathbf{P}_{[t-1]})\right]$. Then, $\Delta p_{j,[t]}$ is denoted as
\begin{equation}
\Delta p_{j,[t]}=\frac{-f'(\mathbf{P}_{[t-1]})+\mathrm{sgn}\left[f'(\mathbf{P}_{[t-1]})\right]\sqrt{\left\{f'(\mathbf{P}_{[t-1]})\right\}^2-2f''(\mathbf{P}_{[t-1]})f(\mathbf{P}_{[t-1]})}}{f''(\mathbf{P}_{[t-1]})}.\label{eq:PowerAlloc_3}
\end{equation}
The partial derivatives $f_{i}(\mathbf{P}_{[t-1]})$, $f_i^{(1)}(\mathbf{P}_{[t-1]})$ and $f_i^{(2)}(\mathbf{P}_{[t-1]})$ can be immediately derived by definition as in \eqref{eq:partial_2}, \eqref{eq:partial_3}, and \eqref{eq:partial_4}, respectively.

\section{Proof of Proposition \ref{prop:Given-remained-data}\label{sec:Appendix_D}}
From the constraint \eqref{eq:P4_c} and \eqref{eq:p_jt}, we have
\begin{equation}
\Delta p_{j,[t]}\leq p_{j,[t-1]}^{(c_{2})}.\label{eq:condition_power_each}
\end{equation}
In transferring transmission power from RB $c_1$ to RB $c_2$, increased power on RB $c_1$ of SBS $j$ increases the SINR of SBS $j$ on RB $c_1$, and reduced power on RB $c_2$ of SBS $j$ increases the SINR of other SBSs on RB $c_2$. Here, the maximum possible $\Delta p_{j,[t]}$ that satisfies the backhaul constraint is obtained in terms of remaining backhaul capacity. The remaining backhaul capacity of SBS $j$ for given $\mathbf{X},\mathbf{P}_{[t]}$ is denoted as
\begin{equation}
    L_j(\mathbf{X},\mathbf{P}_{[t]})=Z_j-\left.\sum_{i\in\mathcal{N}}\sum_{c\in\mathcal{C}}R_{ij}^{(c)}x_{ij}^{(c)}\right|_{\mathbf{P}=\mathbf{P}_{[t]}}.
\end{equation}

i) SINR of SBS $j$ on RB $c_1$

Because the power of SBS $j$ on RB $c_1$ increases, the $\mathrm{SINR}_{ij}^{(c_1)}$ increases. In order to satisfy backhaul constraint \eqref{eq:P3_c},
\begin{equation}
\left.W\log_{2}(1+\mathrm{SINR}_{i^{(j,c_1)}j}^{(c_1)})\right|_{\mathbf{P}=\mathbf{P}_{[t]}}-\left.W\log_{2}(1+\mathrm{SINR}_{i^{(j,c_1)}j}^{(c_1)})\right|_{\mathbf{P}=\mathbf{P}_{[t-1]}}\leq L_{j}(\mathbf{X},\mathbf{P}_{[t-1]}),\label{eq:condition_MyBS_1}
\end{equation}
where $i^{(j,c_1)} = \argmax_{i\in\mathcal{N}}x_{ij}^{(c_1)}$.
The left-hand side of \eqref{eq:condition_MyBS_1} is bounded as follow:
\begin{align}
& \left.W\log_{2}(1+\mathrm{SINR}_{i^{(j,c_1)}j}^{(c_1)})\right|_{\mathbf{P}=\mathbf{P}_{[t]}}-\left.W\log_{2}(1+\mathrm{SINR}_{i^{(j,c_1)}j}^{(c_1)})\right|_{\mathbf{P}=\mathbf{P}_{[t-1]}}\\
&=  W\log_{2}\left(1+X\left(1+\frac{\Delta p_{j,[t]}}{p_{j,[t-1]}^{(c_{1})}}\right)\right)-\log_{2}\left(1+X\right)\\
&\leq  W\log_{2}\left(1+\frac{\Delta p}{p_{j,[t-1]}^{(c_{1})}}\right),\label{eq:condition_myBS}
\end{align}
where $X=\left.\mathrm{SINR}_{i^{(j,c_1)}j}^{(c_1)}\right|_{\mathbf{P}=\mathbf{P}_{[t-1]}}$.
Then, from \eqref{eq:condition_MyBS_1} and \eqref{eq:condition_myBS}, $\Delta p_{j,[t]}$ satisfies the backhaul constraint if
\begin{equation}
    \Delta p_{j,[t]}\leq\left(2^{\frac{L_{j}(\mathbf{X},\mathbf{P}_{[t-1]})}{W}}-1\right)p_{j,[t-1]}^{(c_{1})}.\label{eq:condition_final_myBS}
\end{equation}

ii) SINR of other SBSs on RB $c_2$

The SINR of SBSs except SBS $j$ on RB $c_2$ increases because the interference on RB $c_2$ is decreased. Then, the backhaul constraint \eqref{eq:P3_c} on SBS $k$ is satisfied if
\begin{equation}
\left.{W\log_{2}\left(1+\mathrm{SINR}_{i^{(k,c_2)}k}^{(c_{2})}\right)}\right|_{\mathbf{P}=\mathbf{P}_{[t]}}-\left.{W\log_{2}\left(1+\mathrm{SINR}_{i^{(k,c_2)}k}^{(c_{2})}\right)}\right|_{\mathbf{P}=\mathbf{P}_{[t-1]}}\leq L_{k}(\mathbf{X},\mathbf{P}_{[t-1]}),\label{eq:condition_youBS_1}
\end{equation}
\pagebreak[0]where $i^{(j,c_2)} = \argmax_{i\in\mathcal{N}}x_{ij}^{(c_2)}$.
The left-hand side of \eqref{eq:condition_youBS_1} is bounded by
\pagebreak[0]
\begin{align}
& \left.{W\log_{2}\left(1+\mathrm{SINR}_{i^{(k,c_2)}k}^{(c_{2})}\right)}\right|_{\mathbf{P}=\mathbf{P}_{[t]}}-\left.{W\log_{2}\left(1+\mathrm{SINR}_{i^{(k,c_2)}k}^{(c_{2})}\right)}\right|_{\mathbf{P}=\mathbf{P}_{[t-1]}}\\
&= W\log_{2}\left(1+\frac{\mathrm{SNR}_{i^{(k,c_2)}k}^{(c_2)}p_{k,[t-1]}^{(c_{2})}}{Y-\mathrm{SNR}_{i^{(k,c_2)}j}^{(c_{2})}\Delta p_{j,[t]}}\right)-W\log_{2}\left(1+\frac{\mathrm{SNR}_{i^{(k,c_2)}k}^{(c_{2})}p_{k,[t-1]}^{(c_{2})}}{Y}\right)\\
&= W\log_{2}\left(\frac{M-\mathrm{SNR}_{i^{(k,c_2)}j}^{(c_{2})}\Delta p_{j,[t]}}{Y-\mathrm{SNR}_{i^{(k,c_2)}j}^{(c_{2})}\Delta p_{j,[t]}}\right)-W\log_{2}\left(\frac{M}{Y}\right)\\
&\leq -W\log_{2}\left(1-\frac{\Delta p_{j,[t]}}{p_{j,[t-1]}^{(c_{2})}}\right),
\label{eq:condition_youBS}
\end{align}
where $Y=\sigma^2+\sum_{l\in\mathcal{B}}p_{l,[t-1]}^{(c_{2})}\mathrm{SNR}_{i^{(k,c_2)}l}^{(c_{2})}$, $M=Y+\mathrm{SNR}_{i^{(k,c_2)}k}^{(c_{2})}p_{k,[t-1]}^{(c_{2})}$.
Then, from \eqref{eq:condition_youBS_1} and \eqref{eq:condition_youBS}, we have
\begin{equation}
\Delta p_{j,[t]}\leq\left(1-2^{-\frac{L_{k}(\mathbf{X},\mathbf{P}_{[t-1]})}{W}}\right)p_{j,[t-1]}^{(c_{2})}.\label{eq:condition_you_BS_final}
\end{equation}

From \eqref{eq:condition_power_each}, \eqref{eq:condition_myBS} and \eqref{eq:condition_youBS}, The condition of $\Delta p_{j,[t]}$ that satisfied the backhaul constraint \eqref{eq:P3_c} is denoted in \eqref{eq:limit_delta_p}.
\end{appendices}

\scriptsize{\bibliographystyle{IEEEtran}
\bibliography{IEEEabrv,smallcell}}

\begin{thebibliography}{10}
\providecommand{\url}[1]{#1}
\csname url@samestyle\endcsname
\providecommand{\newblock}{\relax}
\providecommand{\bibinfo}[2]{#2}
\providecommand{\BIBentrySTDinterwordspacing}{\spaceskip=0pt\relax}
\providecommand{\BIBentryALTinterwordstretchfactor}{4}
\providecommand{\BIBentryALTinterwordspacing}{\spaceskip=\fontdimen2\font plus
\BIBentryALTinterwordstretchfactor\fontdimen3\font minus
  \fontdimen4\font\relax}
\providecommand{\BIBforeignlanguage}[2]{{%
\expandafter\ifx\csname l@#1\endcsname\relax
\typeout{** WARNING: IEEEtran.bst: No hyphenation pattern has been}%
\typeout{** loaded for the language `#1'. Using the pattern for}%
\typeout{** the default language instead.}%
\else
\language=\csname l@#1\endcsname
\fi
#2}}
\providecommand{\BIBdecl}{\relax}
\BIBdecl

\bibitem{A_Gupta_WCL15}
A.~K. Gupta, X.~Zhang, and J.~G. Andrews, ``{SINR} and throughput scaling in
  ultradense urban cellular networks,'' \emph{IEEE Wireless Communications
  Letters}, vol.~4, no.~6, pp. 605--608, Dec. 2015.

\bibitem{M_T.Kawser_JEE12}
M.~T. Kawser, H.~M. Farid, A.~R. Hasin, A.~M.~J. Sadik, and I.~K. Razu,
  ``{P}erformance {C}omparison between {R}ound {R}obin and {P}roportional
  {F}air {S}cheduling {M}ethods for {LTE},'' \emph{International Journal of
  Information and Electroniscs Enginerring}, vol.~2, no.~5, pp. 678--681, Sept.
  2012.

\bibitem{Q_Kuang16_TSP}
Q.~Kuang, W.~Utschick, and A.~Dotzler, ``Optimal joint user association and
  multi-pattern resource allocation in heterogeneous networks,'' \emph{IEEE
  Transactions on Signal Processing}, vol.~64, no.~13, pp. 3388--3401, July
  2016.

\bibitem{H_ju13_WCL}
H.~Ju, B.~Liang, J.~Li, and X.~Yang, ``Dynamic power allocation for throughput
  utility maximization in interference-limited networks,'' \emph{IEEE Wireless
  Communications Letters}, vol.~2, no.~1, pp. 22--25, Feb. 2013.

\bibitem{S_Han_PIMRC12}
S.~Han, B.~H. Soong, and Q.~D. La, ``Subcarrier allocation in multi-cell ofdma
  wireless networks with non-coherent base station cooperation and controllable
  fairness,'' in \emph{Proc.2012 IEEE 23rd International Symposium on Personal,
  Indoor and Mobile Radio Communications - (PIMRC)}, Sept. 2012, pp. 524--529.

\bibitem{Z_shen_GLC03}
Z.~Shen, J.~G. Andrews, and B.~L. Evans, ``Optimal power allocation in
  multiuser ofdm systems,'' in \emph{Proc.Global Telecommunications Conference,
  2003. GLOBECOM '03. IEEE}, vol.~1, Dec. 2003, pp. 337--341 Vol.1.

\bibitem{Y_Zhang_TWC04}
Y.~J. Zhang and K.~B. Letaief, ``Multiuser adaptive subcarrier-and-bit
  allocation with adaptive cell selection for ofdm systems,'' \emph{IEEE
  Transactions on Wireless Communications}, vol.~3, no.~5, pp. 1566--1575,
  Sept. 2004.

\bibitem{N_Ksairi_TSP10}
N.~Ksairi, P.~Bianchi, P.~Ciblat, and W.~Hachem, ``Resource allocation for
  downlink cellular ofdma systems -- part i: Optimal allocation,'' \emph{IEEE
  Transactions on Signal Processing}, vol.~58, no.~2, pp. 720--734, Feb. 2010.

\bibitem{N_Ksairi_TSP10_2}
------, ``Resource allocation for downlink cellular ofdma systems -- part ii:
  Practical algorithms and optimal reuse factor,'' \emph{IEEE Transactions on
  Signal Processing}, vol.~58, no.~2, pp. 735--749, Feb. 2010.

\bibitem{D_Fooladivanda13_TWC}
D.~Fooladivanda and C.~Rosenberg, ``Joint resource allocation and user
  association for heterogeneous wireless cellular networks,'' \emph{IEEE
  Transactions on Wireless Communications}, vol.~12, no.~1, pp. 1368--1373,
  Jan. 2013.

\bibitem{Derrick_TWC}
D.~W.~K. Ng, E.~S. Lo, and R.~Schober, ``{Energy-efficient resource allocation
  in multi-cell OFDMA systems with limited backhaul capacity},'' \emph{IEEE
  Transactions on Wireless Communications}, vol.~11, no.~10, pp. 3618 -- 3631,
  Oct. 2012.

\bibitem{Ghime_JSAC}
J.~Ghimire and C.~Rosenberg, ``Revisiting scheduling in heterogeneous networks
  when the backhaul is limited,'' \emph{IEEE Journal on Selected Areas in
  Communications}, vol.~33, no.~10, pp. 2039--2051, Oct. 2015.

\bibitem{Kaiming_JSAC}
K.~Shen and W.~Yu, ``{Distributed pricing-based user association for downlink
  heterogeneous cellular networks},'' \emph{IEEE Journal on Selected Areas in
  Communications}, vol.~32, no.~6, pp. 1100--1113, June 2014.

\bibitem{Q_Han15}
Q.~Han, B.~Yang, G.~Miao, C.~Chen, X.~Wang, and X.~Guan, ``{Backhaul-Aware User
  Association and Resource Allocation for Energy-Constrained HetNets},''
  \emph{IEEE Transactions on Vehicular Technology}, vol.~PP, no.~99, Mar. 2016.

\bibitem{Q_Ye_TWC}
Q.~Ye, B.~Rong, Y.~Chen, M.~Al-Shalash, C.~Caramanis, and J.~G. Andrews, ``User
  association for load balancing in heterogeneous cellular networks,''
  \emph{IEEE Transactions on Wireless Communications}, vol.~12, no.~6, pp.
  2706--2716, June 2013.

\bibitem{T_Bu_INFO}
T.~Bu, L.~Li, and R.~Ramjee, ``{G}eneralized proportional fair scheduling in
  third generation wireless data networks,'' \emph{{Proceedings IEEE INFOCOM
  2006. 25TH IEEE International Conference on Computer Communications}}, pp.
  1--12, Apr. 2006.

\bibitem{T_Girici_JCN}
T.~Girici, C.~Zhu, J.~R. Agre, and A.~Ephremides, ``{P}roportional fair
  scheduling algorithm in {OFDMA}-based wireless systems with {QoS}
  constraints,'' \emph{Journal of Communications and Networks}, vol.~12, no.~1,
  pp. 30--42, Feb. 2010.

\bibitem{Y.C.Chen}
Y.~C. Chen and H.~Y. Hsieh, ``{Joint resource allocation and power control for
  CoMP transmissions in LTE-A HetNets with RRHs},'' \emph{2014 IEEE Wireless
  Communications and Networking Conference (WCNC)}, pp. 1368--1373, Apr. 2014.

\bibitem{A_Abdelnasser15_TCM}
A.~Abdelnasser, E.~Hossain, and D.~I. Kim, ``Tier-aware resource allocation in
  ofdma macrocell-small cell networks,'' \emph{IEEE Transactions on
  Communications}, vol.~63, no.~3, pp. 695--710, Mar. 2015.

\bibitem{K_son_TWC}
K.~Son, S.~Chong, and G.~D. Veciana, ``{D}ynamic association for load balancing
  and interference avoidance in multi-cell networks,'' \emph{IEEE Transactions
  on Wireless Communications}, vol.~8, no.~7, pp. 3566--3576, July 2009.

\bibitem{S.Maghsudi_16}
S.~Maghsudi and E.~Hossain, ``{Distributed downlink user association in small
  cell networks with energy harvesting},'' in \emph{Proc.2016 IEEE
  International Conference on Communications (ICC)}, May 2016, pp. 1 -- 6.

\bibitem{The_Future_of_WN}
M.Guizani, H.~Chen, and C.~Wang, \emph{The Future of Wireless Networks :
  Architectures, Protocols, and Services}.\hskip 1em plus 0.5em minus
  0.4em\relax CRC Press, 2005.

\bibitem{D.Bethanabhotla_14}
D.~Bethanabhotla, O.~Y. Bursalioglu, H.~C. Papadopoulos, and G.~Caire, ``{User
  association and load balancing for cellular massive MIMO},'' in \emph{Proc.
  Information Theory and Applications Workshop (ITA), 2014}, Feb. 2014, pp. 1
  -- 10.

\bibitem{Z_Cui14_CISS}
Z.~Cui and R.~Adve, ``Joint user association and resource allocation in small
  cell networks with backhaul constraints,'' in \emph{Proc. 48th Annual
  Conference on Information Sciences and Systems (CISS)}, Mar. 2014, pp. 1--6.

\bibitem{DBLP_17}
D.~Bethanabhotla, O.~Y. Bursalioglu, H.~C. Papadopoulos, and G.~Caire,
  ``Optimal user-cell association for massive mimo wireless networks,''
  \emph{IEEE Transactions on Wireless Communications}, vol.~15, no.~3, pp.
  1835--1850, Mar. 2016.

\bibitem{3gpp_36_872}
\BIBentryALTinterwordspacing
3GPP, ``Small cell enhancements for {E-UTRA} and {E-UTRAN} - physical layer
  aspects,'' 3rd Generation Partnership Project (3GPP), 3GPP, Valbonne, France,
  Tech. Rep. TR 36.872 v12.1.0, TR 36.872, Dec. 2013. [Online]. Available:
  \url{www.3gpp.org/dynareport/36872.htm}
\BIBentrySTDinterwordspacing

\bibitem{Y_Xia_ring_sol}
Y.~Xia, ``{N}ew optimality conditions for quadratic optiization problems with
  binary constraints,'' \emph{Optimization Letters}, vol.~3, pp. 253--263, Mar.
  2009.

\bibitem{3gpp_36_874}
\BIBentryALTinterwordspacing
3GPP, ``{Coordinated multi-point operation for LTE with non-ideal backhaul},''
  {3rd Generation Partnership Project (3GPP)}, TR {36.874}, Dec. 2013.
  [Online]. Available: \url{www.3gpp.org/dynareport/36874.htm}
\BIBentrySTDinterwordspacing

\end{thebibliography}

\end{document}